\documentclass[useAMS,usenatbib]{mn2e}
\usepackage{amssymb}
\usepackage{graphicx}

\title[Modelling Shear Flows with VINE and FLASH]
      {Modelling Shear Flows with SPH and Grid Based Methods}
      \author[Junk et al.]
{\parbox[t]{\textwidth}{Veronika Junk$^1$\thanks{vjunk@usm.uni-muenchen.de}, 
Stefanie Walch$^{2}$, 
Fabian Heitsch$^3$,
Andreas Burkert$^1$,
Markus Wetzstein$^4$, 
Marc Schartmann$^{1,5}$
Daniel Price$^6$}\vspace*{3pt}\\
$^{1}$University Observatory Munich, Scheinerstrasse 1, 81679 Munich, Germany \\
$^{2}$School of Physics and Astronomy, University of Cardiff, The
Parade, Roath, Cardiff, Cardiff, CF24, United Kingdom \\
$^{3}$Department of Astronomy, University of Michigan, 500 Church St,
Ann Arbor, MI 48109-1042, USA \\
$^{4}$Department of Astrophysical Sciences, Princeton
University,Princeton, NJ 08544, USA \\
$^{5}$Max Planck Institute for Extraterrestrial Physics,
Giessenbachstrasse, 85748 Garching, Germany\\
$^{6}$School of Mathematical Sciences, Monash University, Clayton,Vic 3168, Australia}
\begin{document}

\date{Accepted--- Received---}

\pagerange{\pageref{firstpage}--\pageref{lastpage}} \pubyear{2010}

\maketitle

\label{firstpage}

\begin{abstract}
Given the importance of shear flows for astrophysical gas dynamics, we study the
evolution of the Kelvin-Helmholtz instability (KHI) analytically and numerically. We derive
the dispersion relation for the two-dimensional KHI including viscous dissipation. The
resulting expression for the growth rate is then used to estimate the intrinsic viscosity
of four numerical schemes depending on code-specific as well as on
physical parameters. 
Our set of numerical schemes includes the Tree-SPH code VINE, an alternative SPH
formulation developed by Price (2008), and the finite-volume grid codes
FLASH and PLUTO. 
In the first part, we explicitly demonstrate the effect of dissipation-inhibiting mechanisms such as
the Balsara viscosity on the evolution of the KHI. 
With VINE, increasing density contrasts lead to a continuously
increasing suppression of the KHI (with complete suppression from a
contrast of 6:1 or higher). The alternative SPH formulation including
an artificial thermal conductivity reproduces the analytically expected growth rates
up to a density contrast of 10:1. 
The second part addresses the shear flow evolution with FLASH and
PLUTO. Both codes result in a consistent non-viscous evolution (in the equal as well as in 
the different density case) in agreement with the analytical prediction. 
The viscous evolution studied with FLASH shows minor deviations 
from the analytical prediction. 
\end{abstract}

\begin{keywords}
hydrodynamics - instabilities - methods:analytical - methods: numerical\\
ISM:kinematics and dynamics 
\end{keywords}
\section{Introduction}
Shear flows are an integral part of many astrophysical processes, from jets, the formation of cold streams, to outflows 
of protostars \citep{Dekel_Birnboim_2009, Agertz_2009, Diemand_2008, Walch_Burkert_Naab_Gritschneder_2009}, 
and cold gas clouds falling through the diffuse hot gas in dark matter halos \citep{Bland-Hawthorn_2007,Burkert_2008}. 
Jets and outflows of young stars can entrain ambient material, leading to 
mixing and possibly the generation of turbulence in e.g. molecular clouds \citep{Burkert_2006, Banerjee_Klessen_Fendt_2007, 
Gritschneder_a, Carroll_Frank_Blackman_Cunningham_Quillen_2009},
while the dynamical interaction of cold gas clouds with the background galactic halo
medium can lead to gas stripping of e.g. dwarf spheroidals (e.g. Grcevich \& Putman 2009),
and the disruption of high-velocity clouds
\citep{Quilis_Moore_2001,Heitsch_Putman_2009}. 
The KHI is believed to significantly influence the gas dynamics in all
of these different scenarios. \\
Moreover, viscous flows play a crucial role in e.g. gas accretion onto
galactic discs (\citealp{Das_2008,Park_2009,Heinzeller_Duschl_2009}), as
well as in dissipative processes 
like the turbulent cascade. Typically, the gas viscosity seems to be
rather low in the interstellar medium, with typical flow Reynolds
numbers of $10^5$. \\
To describe these complex processes in detail,
numerical schemes are applied to 
follow the hydrodynamical evolution. 
Numerous simulations use smoothed-particle hydrodynamics (SPH), 
(\citealp{Gingold_Monaghan_1977,Lucy_1977,Review_Benz_1990,Review_Monaghan_1992,Review_Monaghan_2005}),
because its Lagrangian approach allows us to follow the evolution to high densities and small spatial scales. 
In combination with N-body codes, it is a perfect tool for
cosmological simulations (e.g. \citealp{Hernquist_Katz_1989, Couchmann_1995, Springel_2002, 
Marri_2003, Serna_2003}) and galaxy formation and evolution \citep{Katz_1992, 
Evrard_1994, Navarro_1995, Steinmetz_1999, Thacker_2000,
Steinmetz_2002, Naab_Jesseit_Burkert_2006}. 
SPH describes the physical properties of a fluid by smoothing over a representative
set of particles. However, this can lead to several problems. 
It can fail to correctly model sharp density gradients such
as contact discontinuities, or velocity gradients occurring in
e.g. shear flows (see \citealp{Agertz_2007}), thus suppressing shear instabilities such as the KHI. \\
An interesting problem to test the limitations of SPH as well as grid
codes is the passage of a cold dense gas
cloud moving through a hot and less dense ambient medium 
\citep{Murray_1993, Vietri_Ferrara_Miniati_1997, Agertz_2007}.
Such a configuration would
be typical for gas clouds raining onto galactic protodisks, for
High-Velocity Clouds in the Milky Way and for cold HI clouds in the
Galactic disk. 
\citet{Murray_1993} demonstrated using a grid code that in the absence of thermal instabilities and/or
gravity clouds moving through a diffuse gas should be disrupted by hydrodynamical shear flow instabilities within the
time they need to travel through their own mass. 
Agertz et al. (2007) have shown that the KHI, and therefore the disintegration of such clouds is
suppressed in SPH simulations. 
This problem, in particular the suppression of the KHI, has been subject to 
recent discussion in the literature. Several solutions have been proposed, e.g.  
\cite{Price_2007} discusses a mechanism, which involves 
a special diffusion term (see also \citealp{Wadsley_2008}). \\
Furthermore, \cite{Read_Hayfield_Agertz_2009} identify two effects
occurring in the SPH formalism, 
each one separately contributing to the instability suppression.
The first problem is related to the leading order error in the momentum equation, which should decrease 
with increasing neighbor number. However, numerical instabilities prevent 
its decline. By introducing appropriate kernels, 
\cite{Read_Hayfield_Agertz_2009} showed that this problem can be cured.
The second problem arises due to the entropy conservation. 
Entropy conservation inhibits particle mixing and leads to a pressure discontinuity. 
This can be avoided by using a temperature weighted density following
\cite{Ritchie_2001}. 
Recently, \cite{Abel_2010} has shown to solve
this problem by evaluating the pressure force with respect to the local
pressure. In contrast to standard SPH schemes this applies forces to particles only if there is a net force 
acting upon them.
\\
Another characteristic of SPH is the implementation of an artificial
viscosity (AV) term \citep{Monaghan_Gingold_1983}, which 
is necessary in order to treat shock phenomena and to prevent particle
interpenetration.
AV can produce an artificial viscous dissipation in a flow
corresponding to a decrease of the Reynolds-number and
therefore a suppression of the KHI \citep{Review_Monaghan_2005}. 
To confine this effect, a reduction of viscous 
dissipation was proposed by \citet{Balsara_1995} and improved by \citet{Colagrossi_2004}.
\citet{Thacker_2_2000} studied different AV-implementations in SPH and pointed out that the actual choice of
the AV-implementation is the primary factor in
determining code performance. 
An extension of SPH
which includes physical fluid viscosities was discussed by
e.g. \citet{Takeda_1994}, \citet{Flebbe_1994}, \citet{er03}, \citet{Sijacki_Springel_2006} and \citet{Lanzafame_2006}.\\
An alternative to conventional numerical schemes may arise from 
a new class of hybrid schemes based on unstructured grids and combining
the strengths of SPH and grid codes \citep{Springel_2009}. 
Some of the problems listed above might be solved with this type of implementation.\\\\
    In this paper we determine how accurate shear flows and
    the corresponding incompressible KHI are described in common
    numerical schemes. 
    Therefore, in \S~\ref{analytics}, we analytically derive the growth rates of the
    KHI including viscosity. In \S~\ref{numdesc} we briefly describe the numerical
    schemes and outline how 
    the simulations have been analyzed. 
    We then discuss our results.  At first, we concentrate on 
    the standard SPH implementation, which does not contain a physical viscosity
    but instead uses AV. 
    However, as mentioned above, AV does influence the evolution of the flow. 
    In \S~\ref{SPH_simulations}, we discuss the
    ability of two numerical SPH-schemes to model the incompressible KHI,
    namely the Tree-SPH method VINE \citep{Wetzstein_Nelson_Naab_Burkert_2008,
      Nelson_Wetzstein_Naab_2008}, and the SPH code of
    \cite{Price_2007}.\\ 
    By comparing to the derived analytical solution, we asses 
    the effects of AV in VINE and estimate the intrinsic physical
    viscosity caused by AV (\ref{SPH_simulations_equal_density}). 
    We then study the development of the KHI for different
    density contrasts (\ref{SPH_diff_dens_layers}). We show that the instability is suppressed for
    density contrasts equal to or larger than $6:1$. We also discuss the
    remedy suggested by \cite{Price_2007}, hereafter P08.\\
    In \S\ref{Grid_simulations} we then study the same problem with
    two grid codes, FLASH \citep{Fryxell_2000} and
    PLUTO \citep{Mignone_2007}.  As the intrinsic artificial viscosity
    is negligible in these schemes, we study the non-viscous as well
    as the viscous evolution of the KHI for equal (\ref{FLASH_equal}) as well as
    non-equal (\ref{FLASH_diff}) density layers. 
    We summarize our findings in \S\ref{conclusions}.
\section[analytics]{KHI -- analytical description}
\label{analytics}
\begin{figure*}
\begin{center}
\includegraphics[width=0.5\textwidth]{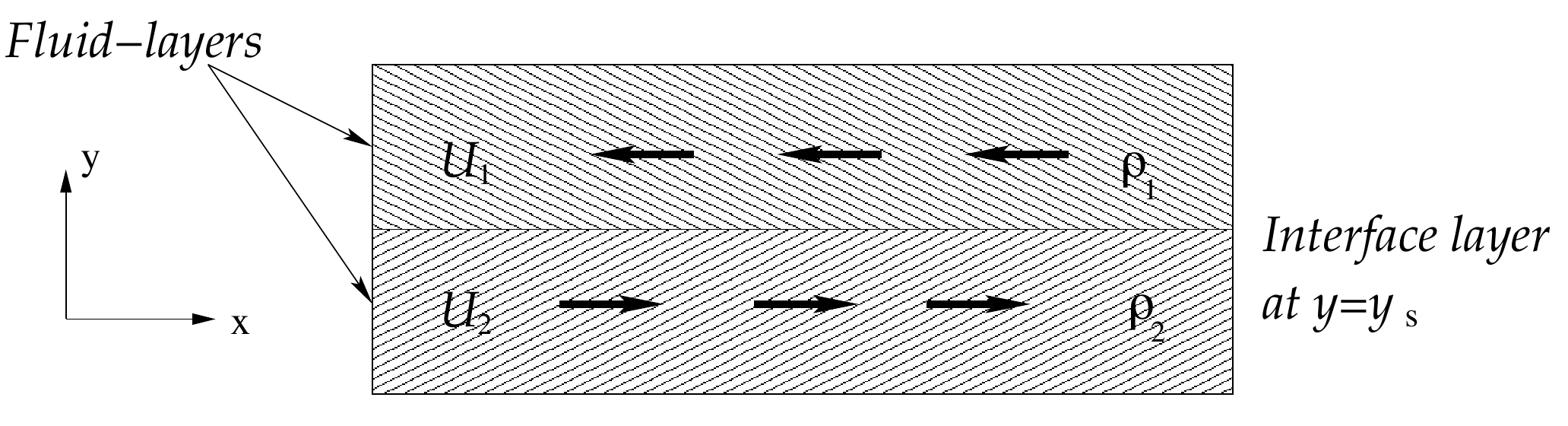}
\caption[Shear flow initial configuration]
{Sketch of the initial conditions considered: 
Two fluid layers with constant densities $\rho_1$ and
  $\rho_2$ flowing in opposite directions 
  with uniform
  velocities $U_1$ and $U_2$.}
\label{Shear_Flow}
\end{center}
\end{figure*}
To derive the growth rate of the KHI in two dimensions including viscosity,
we follow the analysis of \citet{Chandrasekhar} (for a related
analysis see also \citealp{Funada_Joseph_2001} 
and \citealp{Kaiser_2005}). The fluid system is assumed to be viscous and 
incompressible. We use Cartesian coordinates in $x$ and $y$, with two
fluids at densities $\rho_1$, $\rho_2$, and velocities $U_1$, $U_2$ moving
anti-parallel along the $x$-axis, separated by an interface layer at
$y=y_s$ (see Fig.~\ref{Shear_Flow}). 
We neglect the effect of self-gravity.
The hydrodynamical equations for such a system are then given by the
continuity equation
\begin{equation} 
\label{continuity}
\frac{\partial}{\partial t}\rho +
{\bf{\nabla}} \cdot
\left(
\rho {\bf{v}}
\right)=0,
\end{equation}
and the momentum equation 
\begin{equation}
\label{navstokes}
\rho \cdot 
\left[
\frac{\partial {\bf{v}}}{\partial t}+
\left(
{\bf{v}} \cdot {\bf{\nabla}}
\right)
{\bf{v}}
\right]=
-{\bf{\nabla}} p 
+\rho \nu \bigtriangleup {\bf{v}},
\end{equation}
with the flow density $\rho$, velocity ${\bf{v}}$, the thermal pressure $p$
and the kinematic viscosity $\nu$.
\subsection{Linear Perturbations}
\label{perturb}
We linearize equations~\ref{continuity}, and \ref{navstokes} with the perturbations
\begin{eqnarray}
{\bf{v}} &\rightarrow& {\bf{v}} + \delta {\bf{v}} = (U(y)+u;w)\\
\rho &\rightarrow& \rho + \delta \rho,\\
p &\rightarrow& p +\delta p.
\end{eqnarray}
$u$, $w$ express the perturbation in the velocity, $\delta \rho$
and $\delta p$ in the density and pressure, respectively.
This yields the system of linearized equations as
\begin{eqnarray}
\rho\partial_{t} u + \rho U
\partial_{x} u + \rho w
\partial_{y} U &=&-\partial_{x} \delta p+
\nu(\rho+\delta \rho)
 \partial^2_{y}U+\nonumber \\
& &\rho\nu (\partial^{2}_{x}+\partial^{2}_{y}) u,\label{vx} \\
\rho\partial_{t} w+
\rho U \partial_{x} w &=&-\partial_{y}\delta p+
\rho\nu(\partial^{2}_{x}+
\partial^{2}_{y}) w,\label{vz}\\
\partial_{t}
\delta \rho + U \partial_{x}\delta \rho &=&
- w \partial_{y} \rho, \label{lincont}\\
\partial_{t}\delta y_{s} + U_{s} \partial_{x} \delta
y_{s} &=& -w (y_{s}),\label{intz}\\
\delta_{x} u + \delta_{y} w &=& 0.\label{divv}
\end{eqnarray}
Eqs.~\ref{vx} and \ref{vz} represent the 
linearized Navier-Stokes equations, 
where the density may change discontinuously at 
the interface positions denoted by $y_s$. 
Eq.~\ref{lincont} is the linearized continuity
equation. In Eq.~\ref{intz} the subscript $s$ distinguishes the value of the quantity 
at $y=y_s$ (the interface layer). The last equation, Eq.~\ref{divv} 
expresses the incompressibility of the fluid. 
With perturbations of the form
\begin{equation}
\label{perturb_quantities}
u,w,\delta \rho, \delta p, \delta y_s \sim exp[i(k_x x +n t)],
\end{equation}
and assuming that the flow is aligned with the perturbation
vector, i.e. $k=k_x$, we arrive at 
\begin{eqnarray}
\label{dispersionsrelation}
&& D
\left \{
\rho (n + k U) (D w) - k \rho (D U)w
\right \} - 
\rho k^2 (n + k U)w = 
\nonumber \\
& &
i D 
\left \{
\rho \nu k^2 (D w)
\right \} -
i D
\left \{
\rho \nu (D^3 w)
\right \} - 
\nonumber \\
& &
D 
\left\{
k \nu (\rho + \delta \rho) (D^2 U)
\right \} + 
i \rho \nu k^2 (D^2 w) - 
i \rho \nu k^4 w,
\end{eqnarray}
where $D\equiv d/dy$. 
The term, $i \rho \nu k^2 (D^2 w)$ in Eq.~\ref{dispersionsrelation} can be replaced
with 
\begin{equation}
i \rho \nu k^2 (D^2 w) = i k^2 D(\rho \nu (Dw)) - i k^2 (D w)(D(\rho \nu)).
\end{equation}
The boundary condition at $y=y_s$ is determined by an integration 
over an infinitesimal element ($y_s-\epsilon \mbox{ to } y_s + \epsilon$), for 
the limit $\epsilon \rightarrow 0$. Please note, that with Eq.~\ref{lincont} 
it follows for $\delta \rho$, 
\begin{equation}
\delta \rho = i\frac{w}{(n+k_x U)} (D \rho).
\end{equation}
After integration, the boundary condition becomes,
\begin{eqnarray}
\label{boundary}
&& \Delta_s 
\left\{
\rho (n+k U)(D w) -\rho k (D U)w
\right\} = \nonumber \\ 
& & i k^2 \Delta_s 
\left\{ \nu 
\rho (D w)
\right\} - 
i \Delta_s 
\left\{ \nu 
\rho (D^3 w)
\right\} - \nonumber \\
& & k \Delta_s
\left\{ \nu 
\rho (D^2 U)
\right\}-
i k \Delta_s
\left\{ \nu
\frac{w}{(n+k U)} (D \rho) (D^2 U)
\right\} + \nonumber \\
& & 
i k^2 \Delta_s
\left\{
\nu \rho (D w)
\right\} -
i k^2 \lim_{\epsilon \rightarrow 0}{
\int_{y_s-\epsilon}^{y_s+\epsilon}{(D w) D(\nu \rho)\,dy}}
\end{eqnarray}
where $\Delta_s$ is specifying the jump of any continuous quantity $f$
at $y=y_s$, 
\begin{equation}
\Delta_s(f)=f_{(y=y_s+0)}-f_{(y=y_s-0)}.
\end{equation}
For $\nu\equiv 0$ we retrieve the corresponding expression as
given by \citet{Chandrasekhar}.
\subsection{Special case: constant velocities and densities}     
To simplify the derivation of the growth rate $n$ further, we 
consider the case of two fluid layers with constant 
densities $\rho_1$ and $\rho_2$, and constant flow velocities $U_1$ 
and $U_2=-U_1$.
In each region of constant $\rho_{1,2}$ and $U_{1,2}$, Eq.~\ref{dispersionsrelation} reduces to,
\begin{eqnarray}
\label{dispersion_constantregion}
& &\left[
(n+kU_{1,2})\rho_{1,2}-2i \nu k^2
\right] (D^2 w) + i \nu (D^4w) - \nonumber \\
& & k^2
\left[
(n+kU_{1,2}) - i \nu k^2
\right]w=0
\end{eqnarray}
The layers are separated at $y=y_s=0$, and  
$w/(n+k U)$ must be continuous at the interface. 
Also, $w$ must be finite for $y\rightarrow \infty$, so that 
the solution of Eq.~\ref{dispersion_constantregion} 
has the following form,
\begin{eqnarray}
w &=& A(n+k U_1)e^{+k y} \mbox{         } (y<0)\\
w &=& A(n+k U_2)e^{-k y} \mbox{         } (y>0).
\end{eqnarray}
We assume that $\nu_1=\nu_2=\nu$ (which is the case if we consider two
media with the same viscous properties). Inserting this in Eq. ~\ref{boundary}, the 
characteristic equation yields, 
\begin{eqnarray}
&& n^2 + 2
\left[
k (\alpha_2 U_2 + \alpha_1 U_1)
-\frac{i k^2 \nu}{2}
\right] n + \\
& & k^2 
\left(
\alpha_2 U^2_2 + \alpha_1 U^2_1
\right) -
i k^3 \nu  
\left(
\alpha_2 U_2 + \alpha_1 U_1
\right)
=0.
\end{eqnarray}
The parameters $\alpha_1$, $\alpha_2$ are defined by,
\begin{eqnarray}
\label{alpha}
\alpha_1 = \frac{\rho_1}{\rho_1+\rho_2}, \mbox{      } 
\alpha_2 = \frac{\rho_2}{\rho_1+\rho_2}.
\end{eqnarray}
\begin{table*}
\label{table_conversion}
\begin{center}
\begin{tabular}{|c|c|c|}\hline
physical parameters & dimensionless &in cgs units \\\hline\hline
Box size & 2 & 2 cm     \\ \hline
Mass & 4 & 2780.81 g \\ \hline
velocity &  0.387  & 0.40 km/s\\ \hline
time     &   1 & 9.8 $\cdot 10^{-6}$s\\ \hline
\end{tabular}
\caption[Conversion of code units to physical units]
{
  Initial conditions in dimensionless units (first column) and in cgs units (second column). In the text we always refer to dimensionless units.
}
\end{center}
\end{table*}
Solving for n, we get the expression for the mode of the linear KHI:
\begin{eqnarray}
\label{KHI_mode}
&& n = - 
\left[
k(\alpha_2 U_2 + \alpha_1 U_1) - \frac{i k^2 \nu}{2}
\right] \pm 
\nonumber \\
& & 
\sqrt{
- k^2 \alpha_1 \alpha_2 (U_1-U_2)^2 -
\frac{k^4 \nu^2}{4}},
\end{eqnarray}
applying $U_2=-U_1=U$ leads to
\begin{eqnarray}
\label{KHI_analytical}
  n &=& 
  \left[
    k^2 U^2 (\alpha_2-\alpha_1) + \frac{i k^2 \nu}{2}
  \right] \pm 
  \nonumber \\
  & & 
  \sqrt{
- 4 k^2 \alpha_1 \alpha_2 U^2 -
\frac{k^4 \nu^2}{4}}.
\end{eqnarray}
The mode is exponentially growing/decaying with time, if 
the square root of $n$ becomes imaginary, 
\begin{eqnarray}
\label{mode_diff_layers}
n &=& 
\left[
k^2 U^2 (\alpha_2-\alpha_1)
\right] + 
\nonumber \\
& &
i
\left[
\frac{\nu k^2}{2} \pm 
\sqrt{\frac{\nu^2 k^4}{4}+4k^2U^2 \alpha_1 \alpha_2
}
\right].
\end{eqnarray}
The first term describes oscillations (which is not of interest for the growth), the second term 
the growth/decay, with a damping due to the viscosity.
We use this formula for the comparison with our 
numerical studies for different density shearing layers. For equal density shearing layers 
$\rho_1=\rho_2=\rho$, Eq.~\ref{mode_diff_layers} leads to
\begin{equation}
\label{slope_equal_densities}
n = i 
\left[
\frac{\nu k^2}{2} \pm  
\left(
\frac{\nu^2 k^4}{4}+ k^2 U^2 
\right)^\frac{1}{2}
\right].
\end{equation}
    In \S~\ref{SPH_simulations} and \S~\ref{Grid_simulations} we use
    the velocity in direction of the perturbation, which in the above analysis refers to the $y$-direction and therefore, 
    to the $v_y$-velocity component ($w$) when comparing with
    simulations. 
    The exponential term in Eq.~\ref{perturb_quantities} ($\sim
    \exp{(i \cdot n \cdot t)}$) describes the time evolution of the
    KHI. In the following, we therefore compare
    $\ln{\left(v_y\right)}$ with the analytical expectation $\ln{(w)}
    \sim i \cdot n \cdot t$.
\section{KHI - numerical description}
\label{numdesc}
We use two independent numerical approaches  - particle based and grid based - to follow the hydrodynamics of the
system. 
In the following, all physical parameters are given in code units (see
table 1 for conversion to physical units). 
\subsection{SPH models - VINE \& P08}
\label{numdescVINE}
The parallel Tree-SPH code VINE
\citep{Wetzstein_Nelson_Naab_Burkert_2008,Nelson_Wetzstein_Naab_2008} 
has been successfully applied to a number of astrophysical problems on
various scales \citep{Naab_Jesseit_Burkert_2006, Jesseit_Naab_Peletier_Burkert_2007,Gritschneder_b,Walch_Burkert_Naab_Gritschneder_2009,Kotarba_2009}. 
In VINE the implementation of AV is based on the description by \cite{Monaghan_Gingold_1983}, 
and it includes the modifications by \cite{Lattanzio_86}. AV is not a
real physical viscosity, but implemented to allow the treatment of
shock phenomena. A viscous term, $\Pi$
\begin{equation}
\Pi = - \nu 
\left(
\frac{{\bf{v}}\cdot {\bf{r}}}{r^2 + \epsilon \bar{h}^2}
\right),
\end{equation}
is added to the SPH momentum equations. The quantity $\epsilon \sim 0.01$ prevents a singularity if
$r \rightarrow 0$, while $\bar{h}$ present the mean smoothing length 
between two particles. For $\nu$ follows,
\begin{equation}
\nu = \frac{\bar{h}}{\bar{\rho}} 
\left(
\alpha \bar{c} - \beta \frac{\bar{h} {\bf{v}} \cdot
  {\bf{r}}}{r^2 + \epsilon \bar{h}^2}
\right),
\end{equation}
$\bar{\rho}$, and $\bar{c}$ are the mean density and the mean sound
speed, respectively.
The AV-parameter $\alpha$ controls the shear and the bulk
viscosity, whereas the $\beta$ parameter regulates the shock-capturing
mechanism. 
In the following we set $\alpha=0.1$, and $\beta=0.2$ if not otherwise specified.
AV reduces the Reynolds-number of the flow,
resulting in the damping of the KHI \citep{Review_Monaghan_2005}. 
\citet{Balsara_1995} proposed a corrective term, improving the behavior of the AV in shear flows.
Further improvements are discussed in \cite{Review_Monaghan_2005} 
and references therein. 
VINE can be run with and without the 'Balsara-viscosity'. \\
To prevent the so-called 'artificial pairing' in SPH (e.g. \citealp{Schuessler_Schmitt_1981}), we implement a
correction developed by \cite{Thomas_Couchman_1992}. 
Details can be found in \cite{Wetzstein_Nelson_Naab_Burkert_2008} and \cite{Nelson_Wetzstein_Naab_2008}. \\
The SPH code presented in P08 uses a different implementation of AV as
explained in \cite{Morris_1997} to prevent the side effects of artificial dissipation. 
Additionally, a diffusion term called 'artificial thermal conductivity' is implemented (see \S~\ref{SPH_diff_dens_layers}), which 
has been shown to prevent the KHI suppression in shear flows with large density
contrasts \citep{Price_2007}.
\subsection{Grid-based models - FLASH \& PLUTO}
\label{numdescFLASH}
We choose the publicly available, MPI-parallel FLASH code version 2.5
\citep{Fryxell_2000}. FLASH is based on the block-structured AMR technique implemented in
the PARAMESH library \citep{MacNeice_2000}. 
However, we do not make use of the AMR refinement technique, but use
uniform grids throughout this paper. 
In FLASH's hydrodynamic module the Navier-Stokes equations are solved
using the piecewise parabolic method \citep{Colella_1984}, which incorporates a Riemann solver to compute fluxes between
individual cells. 
We use a Riemann tolerance value of $10^{-7}$ and a CFL of $0.5$. 
Due to FLASH's hydrodynamic scheme, the intrinsic numerical viscosity
is reduced to a minimum. 
This allows us to study the influence of a physical viscosity on the
growth of the KHI. 
We therefore modify the hydrodynamical equations based on the 
FLASH module 'diffuse' to explicitly include a viscous term, which
scales 
with a given kinematic viscosity (see \ref{FLASH_equal} and \ref{FLASH_diff}). \\
As an additional test, we apply
the Godunov-type high resolution shock capturing scheme PLUTO
\citep{Mignone_2007}. It is a multiphysics, multialgorithm modular
code, especially designed for the treatment of discontinuities. For
the simulations described in this paper, we employ different
Riemann-solvers and time-stepping methods on a uniform, static grid.
\subsection{Initial conditions and analysis method}
\label{initial_conditions_analysis}
Our numerical ICs are identical to the ones used for the derivation
of the analytical growth rates (see \S\ref{analytics},
Fig.~\ref{Shear_Flow} and table 1).
To excite the instability, we apply a velocity perturbation in $y$
direction: 
\begin{equation}
\label{velocity_perturbation}
v_{y} = v_0 \sin(k \cdot x) \cdot \exp 
\left[
-\left(
\frac{y}{\sigma_0}
\right)^2
\right],
\end{equation}
where $k$ is the wavenumber and $v_0$ is the perturbation amplitude of the $y$-velocity triggering the
instability. The parameter $\sigma_0$ controls how quickly the perturbation decreases with $y$
(see discussion Appendix~\ref{dependence_sigma_0}). 
It is set to $\sigma_0=0.1$ if not otherwise specified. 
Initial pressure and density are set to $p_0\equiv 1$ and $\rho_0\equiv 1$, resulting in a
sound speed of $c_{s,0}=\sqrt{5/3}$ with an adiabatic exponent of $\gamma=5/3$. 
Since the analysis of \S\ref{analytics} is only valid for an incompressible
fluid, the flow speed $U$ must be subsonic. We chose $U\equiv 0.3
\times c_{s,0}\approx 0.387$, and the initial perturbation is $v_0=0.1 \times U$. 
We tested the assumption of incompressibility by calculating $\nabla \cdot {\bf{v}}$, which vanishes 
for incompressible flows. 
This is satisfied in the linear regime, the primary focus of our work.
The wavenumber $k$ is equal to $4 \pi/L$, where $L$ is the box length.
The simulated box ranges from $[-1,1]$ in both directions. We use
periodic boundary conditions. If not otherwise specified the AV
parameters are set to $\alpha=0.1$ and $\beta=0.2$. \\\\
To analyze the SPH and grid simulations consistently, we
bin the SPH particles on a $64^2$ grid, using the cloud-in-cell method
\citep{Hockney_Eastwood_1988}.
For the grid codes, the same initial conditions are used. A resolution
of $512^2$ is adopted during the calculation, but we rebin to a $64^2$
grid for the analysis. 
We measure the fastest-growing mode, which is 
the $k=4 \pi/L$ mode of the velocity perturbation in $y$
direction via a Fourier analysis. For more information see
Appendix~\ref{FT_KHI_amplitudes}. \\\\
We perform two sets of simulations with (i) equal 
density layers (see \S~\ref{SPH_simulations_equal_density} for SPH and
\S~\ref{FLASH_equal} for grid codes) and (ii) unequal density layers
(see \S~\ref{SPH_diff_dens_layers} for SPH and \S~\ref{FLASH_diff} for
grid codes). In the latter case we assume pressure equilibrium. For
SPH, we investigate the effects of equal mass and different mass
particles (see \S~\ref{SPH_diff_dens_layers}). 
\begin{figure*}
\begin{center}
\includegraphics[width=0.49\textwidth]{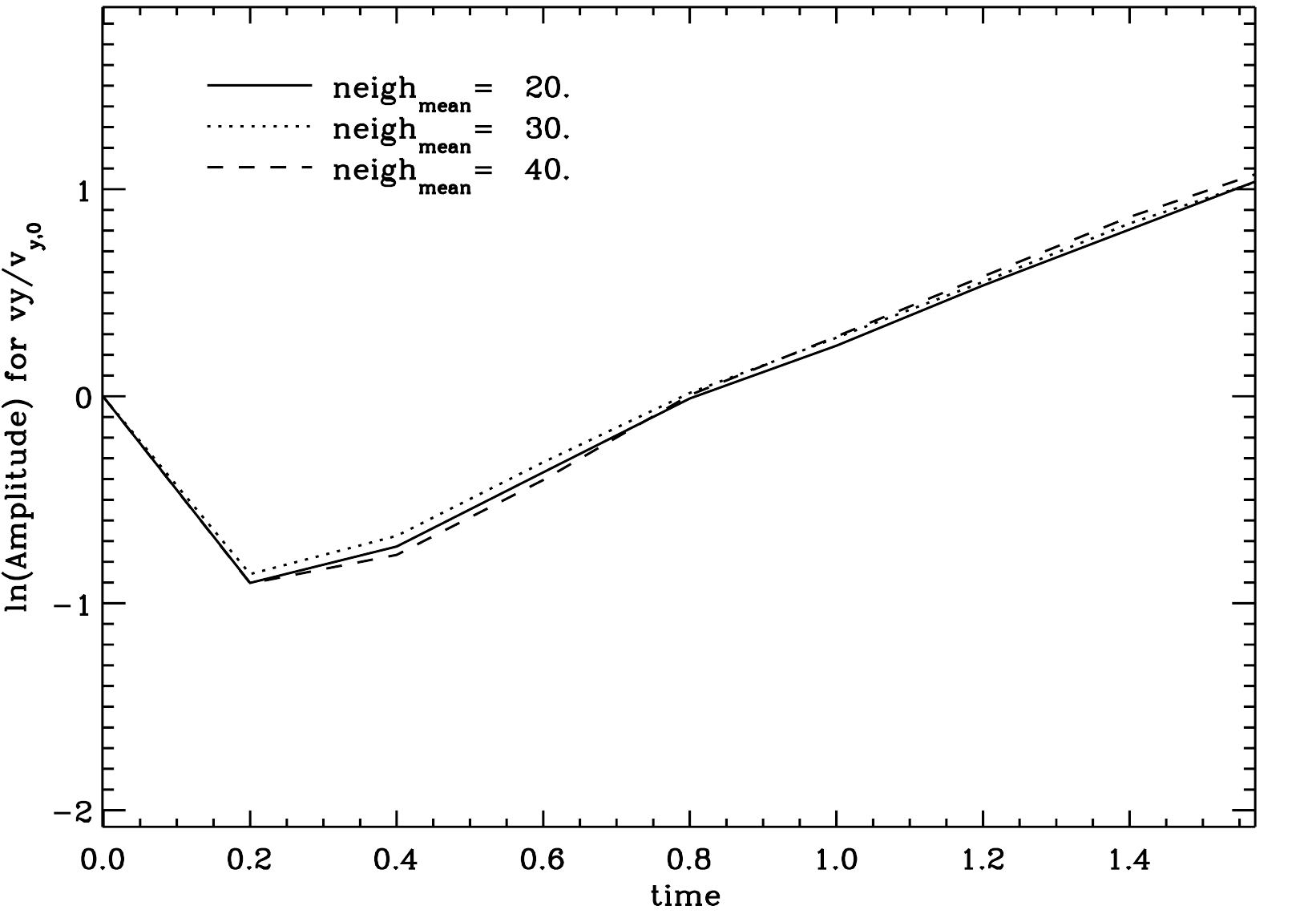}
\includegraphics[width=0.49\textwidth]{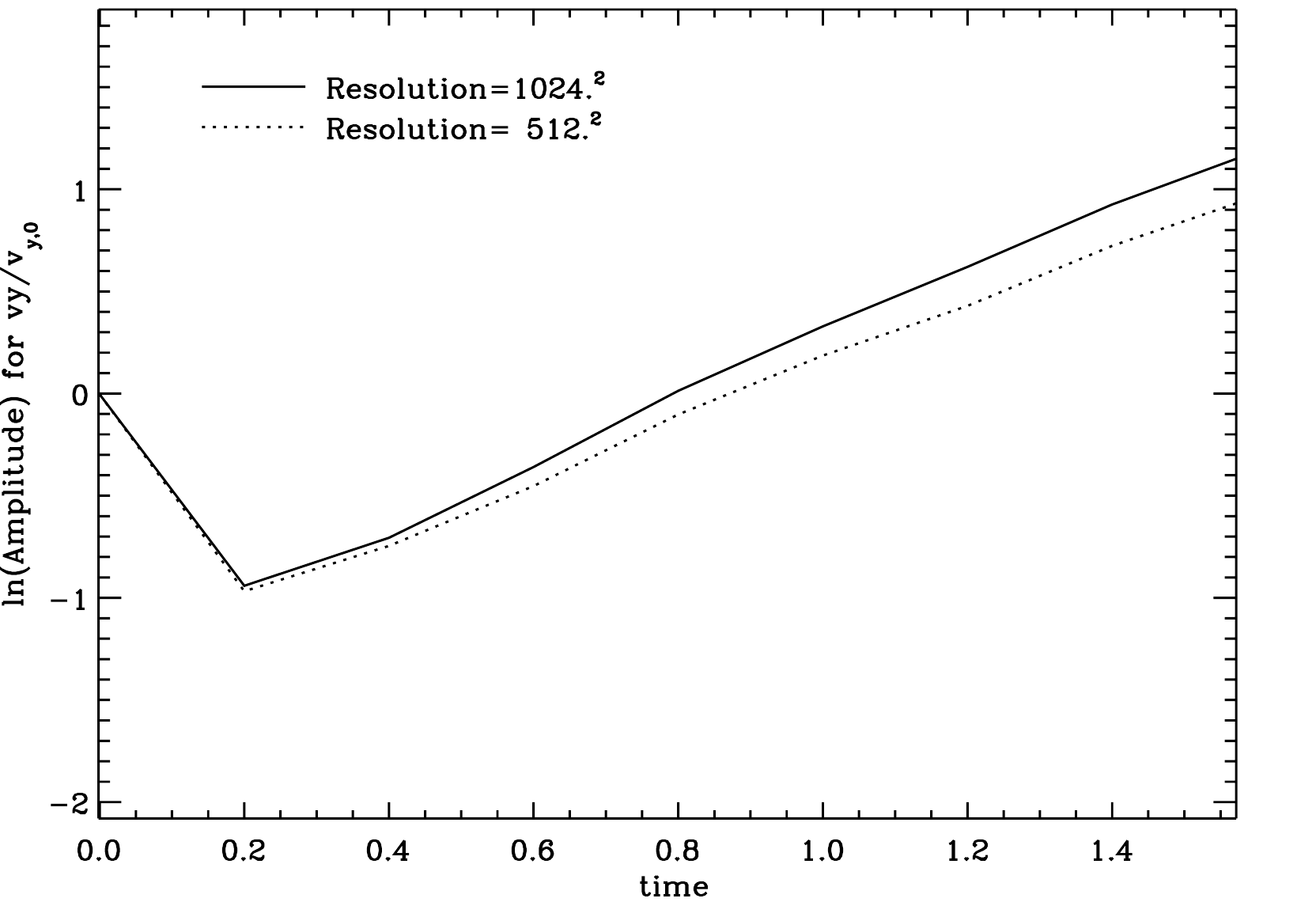}
\caption[]
{
  Time evolution of the $v_y$-amplitude using VINE for different numbers of
  mean neighbors, $\bar{n}_{neigh}$ (left panel, and for different 
  particle number (right panel).
}
\label{multi_equal_neighbors_vy}
\end{center}
\end{figure*}
\section{SPH-Simulations of the KHI}
\label{SPH_simulations}
In the following, we model the evolution of the KHI in systems with $\rho_1=\rho_2$ 
(\S\ref{SPH_simulations_equal_density}) and $\rho_1\neq \rho_2$
(\S\ref{SPH_diff_dens_layers}). 
We apply VINE, if not otherwise specified, and use the analytical
growth rates
(Eqs.~\ref{mode_diff_layers}, and \ref{slope_equal_densities}) derived in
\S~\ref{analytics} to determine the effect of AV.
\subsection{Fluid layers with equal densities:}
\label{SPH_simulations_equal_density}
In the case of $\rho_1=\rho_2$ we vary the following parameters: 
the resolution, which can be either enhanced by using more particles, or decreasing 
the smoothing length $h$, and the AV-parameters $\alpha$ and $\beta$.  We vary one parameter at a time, 
while the other ones are set to the fiducial values (see ~\ref{numdescVINE}).
In the context of AV we discuss the importance of the
Balsara-viscosity. 
In Appendix~\ref{dependence_sigma_0} we also discuss the influence of different $\sigma_0$,
which determines the strength of the initial 
vy-perturbation (Eq.~\ref{velocity_perturbation}).
\begin{figure*}
\begin{center}
\includegraphics[width=0.7\textwidth]{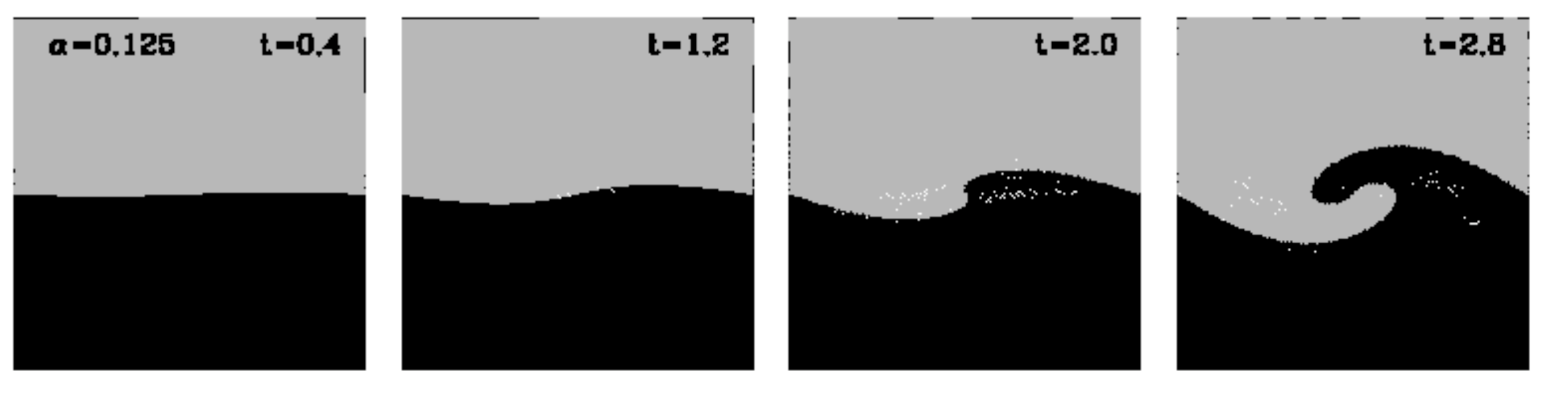}
\includegraphics[width=0.7\textwidth]{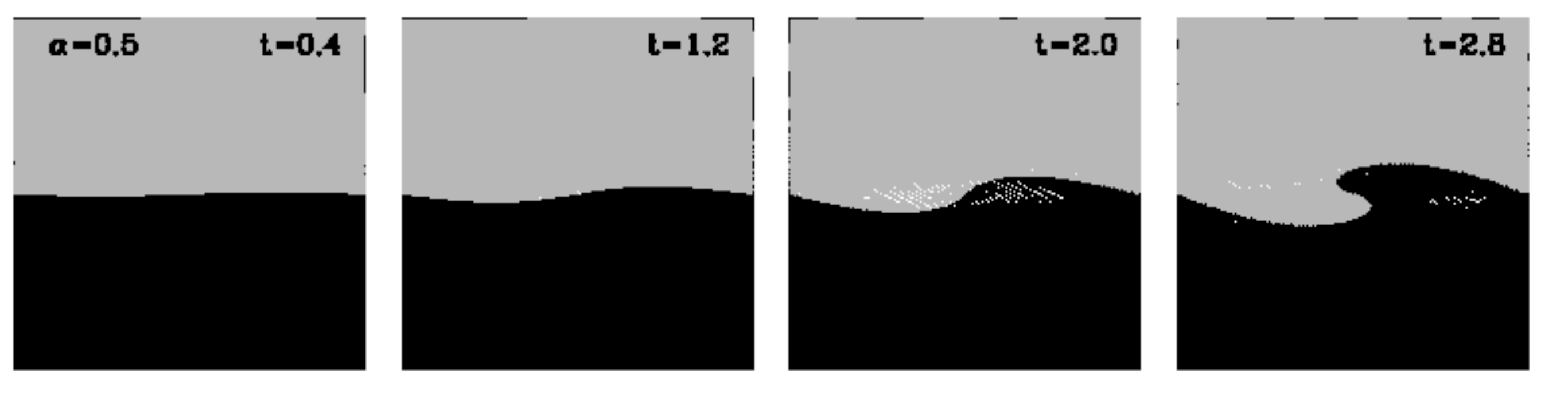}
\includegraphics[width=0.7\textwidth]{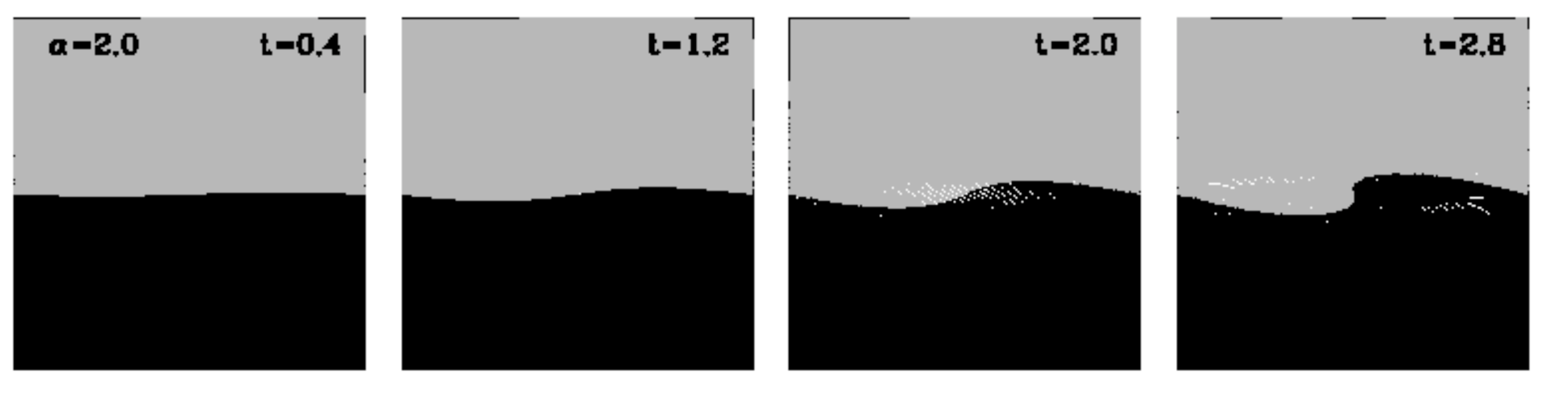}
\caption[Evolution and growth of the KHI for different values for
artificial viscosity parameters $\alpha$]
{
  Time evolution of the KHI using VINE for increasing AV parameter $\alpha$ (top to bottom) and constant $\beta=2$
  The panels show the central region of each simulation box, ranging
  from $[-0.5,0.5]$. The upper layer (grey area) is moving to the left, the lower 
  layer (black area) to the right. Noticeable damping occurs for
  $\alpha > 0.125$ (see left panel of Fig.~\ref{multi_equal_alpha_beta_vy}).
}
\label{Shear_plots_alpha}
\end{center}
\end{figure*}
\begin{figure*}
\begin{center}
\includegraphics[width=0.7\textwidth]{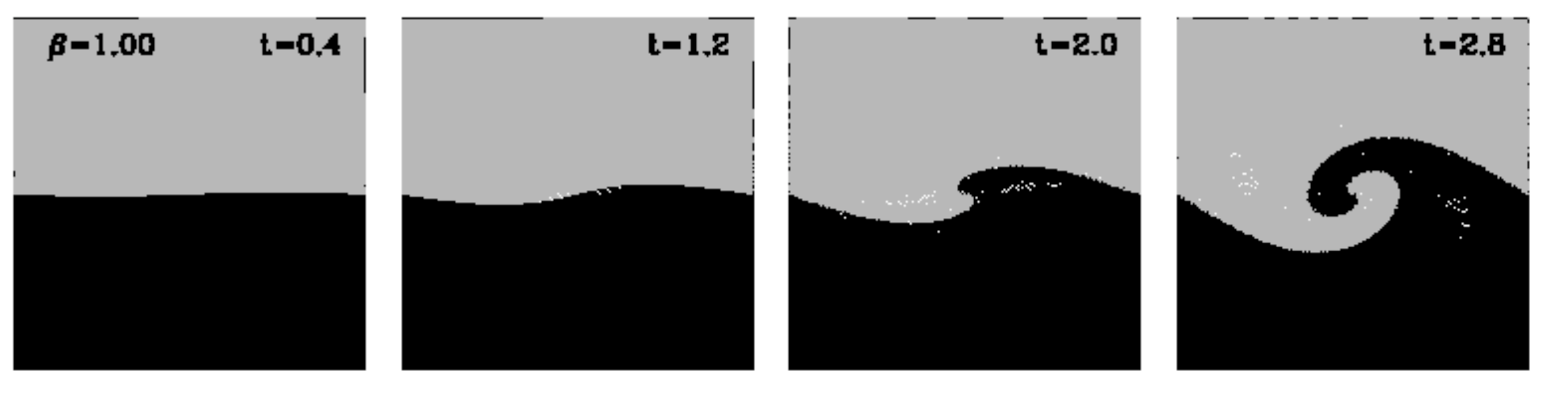}
\includegraphics[width=0.7\textwidth]{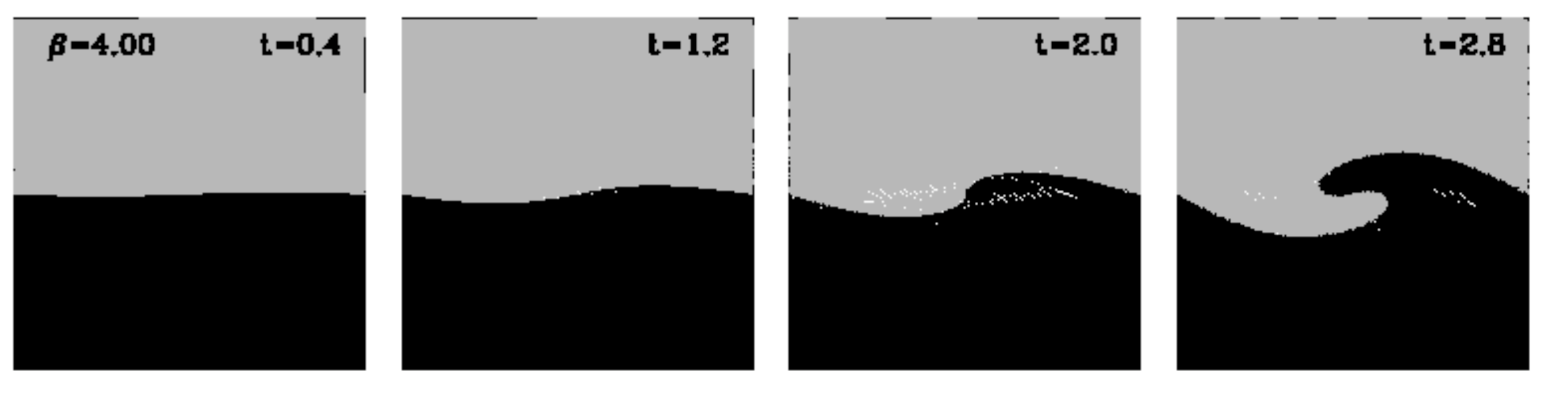}
\includegraphics[width=0.7\textwidth]{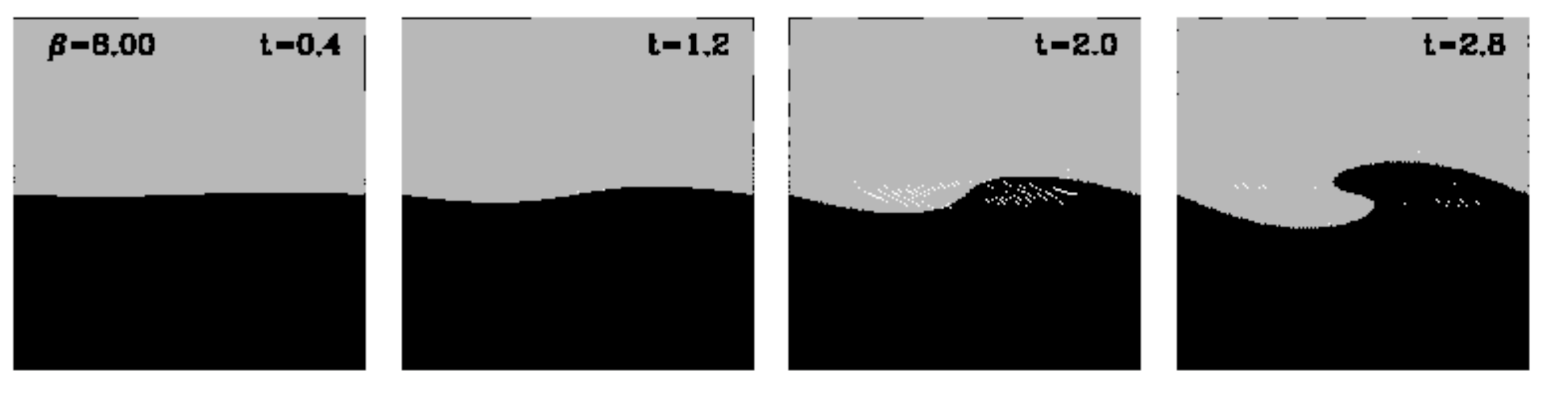}
\caption[Evolution and growth of the KHI for different values for
artificial viscosity parameter $\beta$]
{
  Like Fig.~\ref{Shear_plots_alpha} 
  but for increasing values of the AV-parameter 
  $\beta$ ($\alpha=0.1$). 
  A noticeable damping occurs for of $\beta > 1$ (see right panel of Fig.~\ref{multi_equal_alpha_beta_vy}).
}
\label{Shear_plots_beta}
\end{center}
\end{figure*}
\begin{itemize}
\item 
  {\sc{Dependence on resolution:}} \\
  According to the smoothing procedure in the SPH scheme, each particle requires a certain number of neighboring particles for the 
  calculation of its physical quantities. In VINE, these range from
  $n_{neigh,min}$ to $n_{neigh,max}$. The corresponding mean value
  of neighbors, $\bar{n}_{neigh}$, determines the smoothing length
  $h$. For a constant particle number, increasing $\bar{n}_{neigh}$ leads to a larger smoothing
  length, while at the same time the effective resolution is
  decreased. \\
  In Fig.~\ref{multi_equal_neighbors_vy} we show the time evolution of
  the $v_y$-amplitude, which describes the growth of the KHI. 
  For $t \le 0.2$ the amplitudes decrease since the SPH particles lose
  kinetic energy by moving along the $y$-direction into the area of
  the opposite stream (see Appendix~\ref{dependence_sigma_0}). 
  Therefore we only consider $t> 0.2$ when fitting the growth rates of the KHI. 
  The left panel of Fig.~\ref{multi_equal_neighbors_vy} shows the amplitude growth for 
  $\bar{n}_{neigh}=20$, $30$, and $40$, respectively.  (The commonly
  used value in two dimensions is $\bar{n}_{neigh}=30$). 
  All three cases appear to be similar. 
  Thus, different $\bar{n}_{neigh}$ do not have a substantial impact on the
  KHI-amplitude growth. \\
  The right panel of Fig.~\ref{multi_equal_neighbors_vy} shows
  the dependence on particle number, for the fiducial case of $512^2$ (dotted
  line) and for an increased resolution of $1024^2$ (solid line). 
  The difference for the fitted viscosity is small ($\le 1 \%$).
\end{itemize}
\begin{figure*}
\begin{center}
\includegraphics[width=0.49\textwidth]{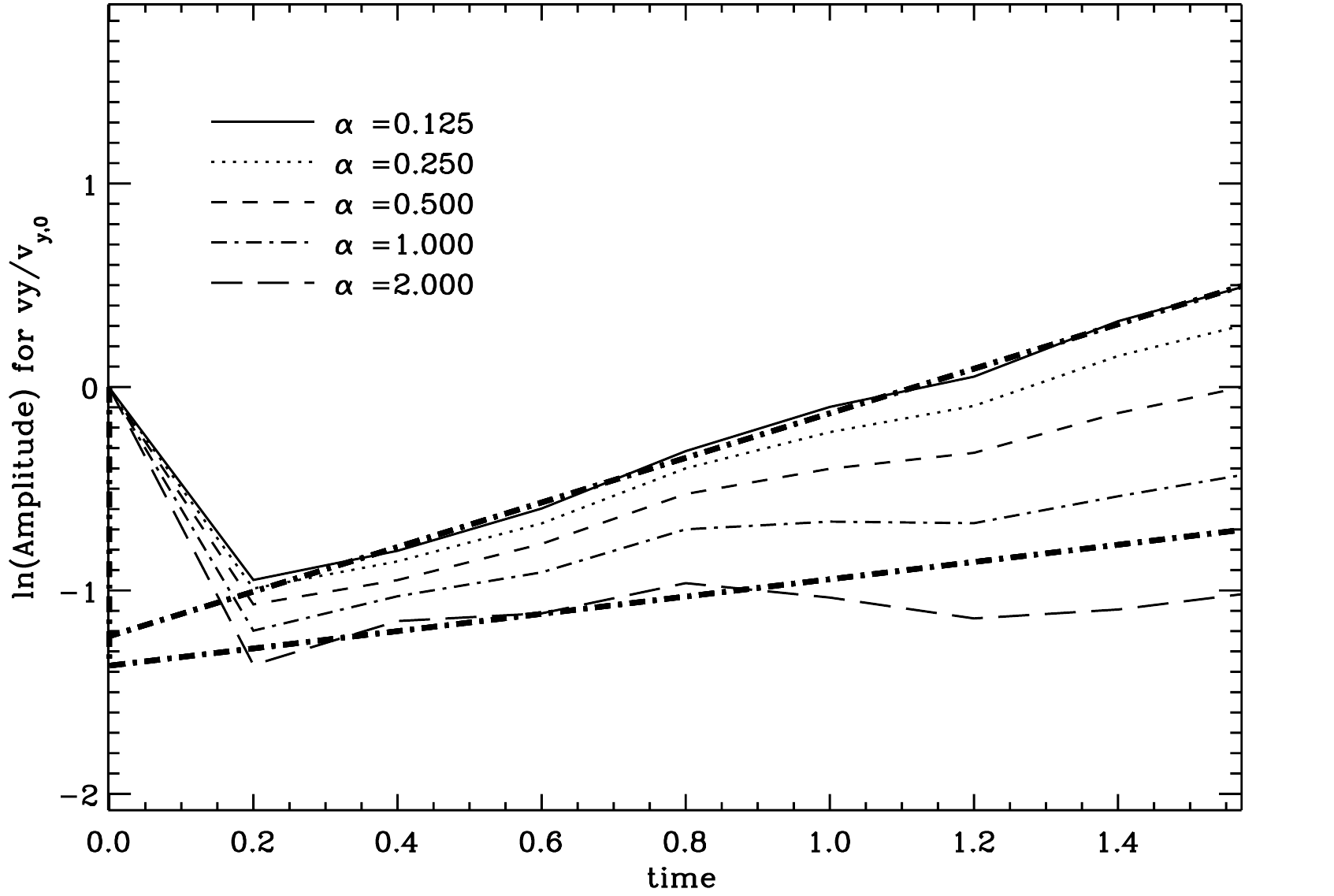}
\hfill
\includegraphics[width=0.49\textwidth]{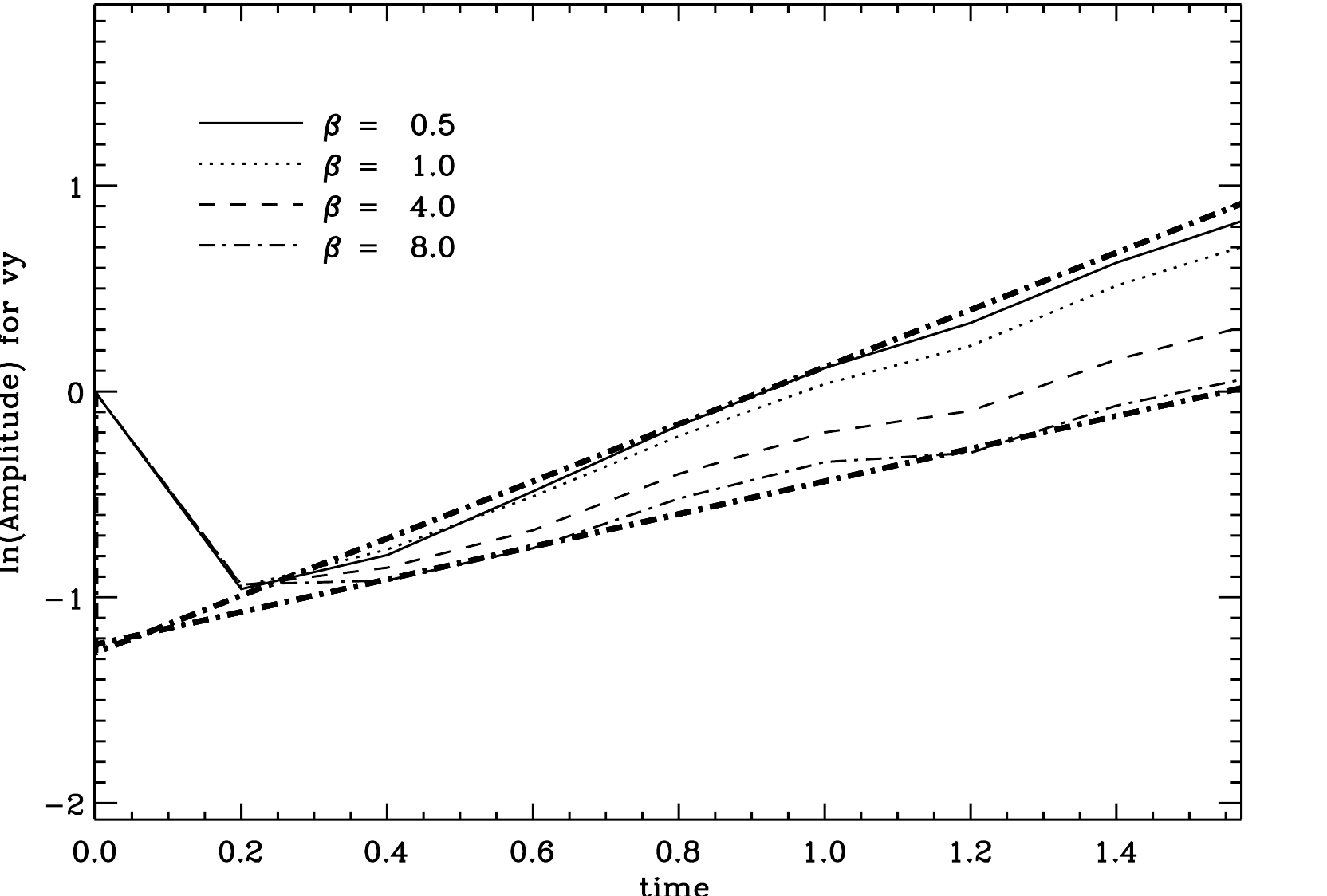}
\caption[]
{
  Left panel: Time evolution of the VINE $v_y$-amplitude for 
  different values of the AV-parameter $\alpha$, where
  $\beta$ has been fixed to $\beta=2$. The
  thick dashed-dotted lines correspond to the analytical fit, shown for
  $\alpha=0.125$ and $\alpha=2$ (which corresponds to
  $\nu_{\mathrm{SPH}}=0.07$ and $\nu_{\mathrm{SPH}}=0.1$). 
  Right panel: Like before, but for different values of the
  AV-parameter $\beta$, where
  $\alpha$ has been fixed to $\alpha=0.1$. 
} 
\label{multi_equal_alpha_beta_vy}
\end{center}
\end{figure*}
\begin{figure*}
\begin{center}
\includegraphics[width=0.49\textwidth]{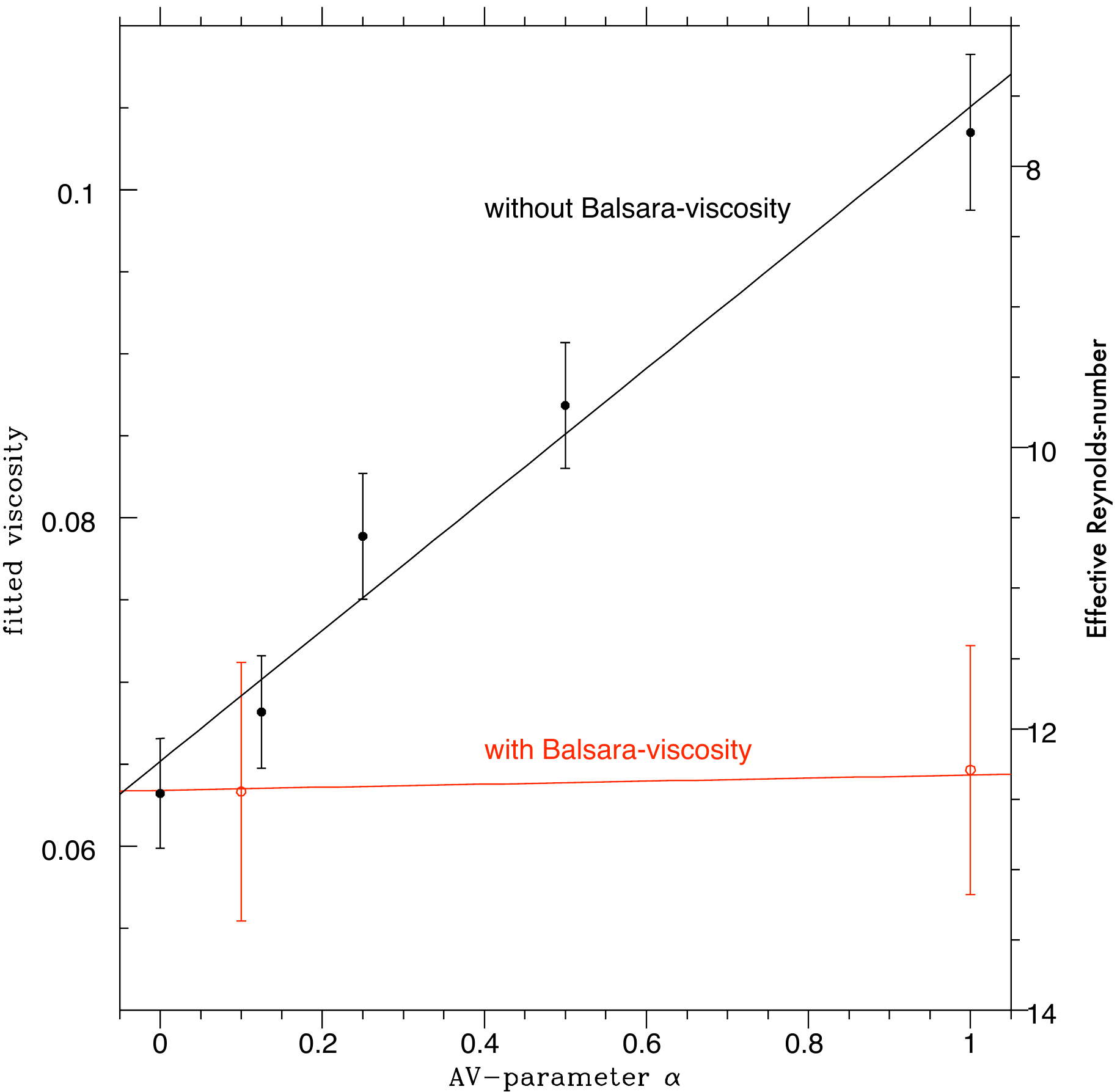}
\caption[VINE, equal density layers: AV-parameter $\alpha$ versus the
fitted viscosity $\nu_{\mathrm{SPH}}$]
{
  Derived physical viscosities ($\nu_{\mathrm{SPH}}$) corresponding to
  different AV parameters $\alpha$ with (open red points) and without
  (filled black points) Balsara-viscosity. We also show the corresponding
  effective $Re$-numbers. 
}
\label{alpha_fit_visc}
\end{center}
\end{figure*}
\begin{figure*}
\begin{center}
\includegraphics[width=0.49\textwidth]{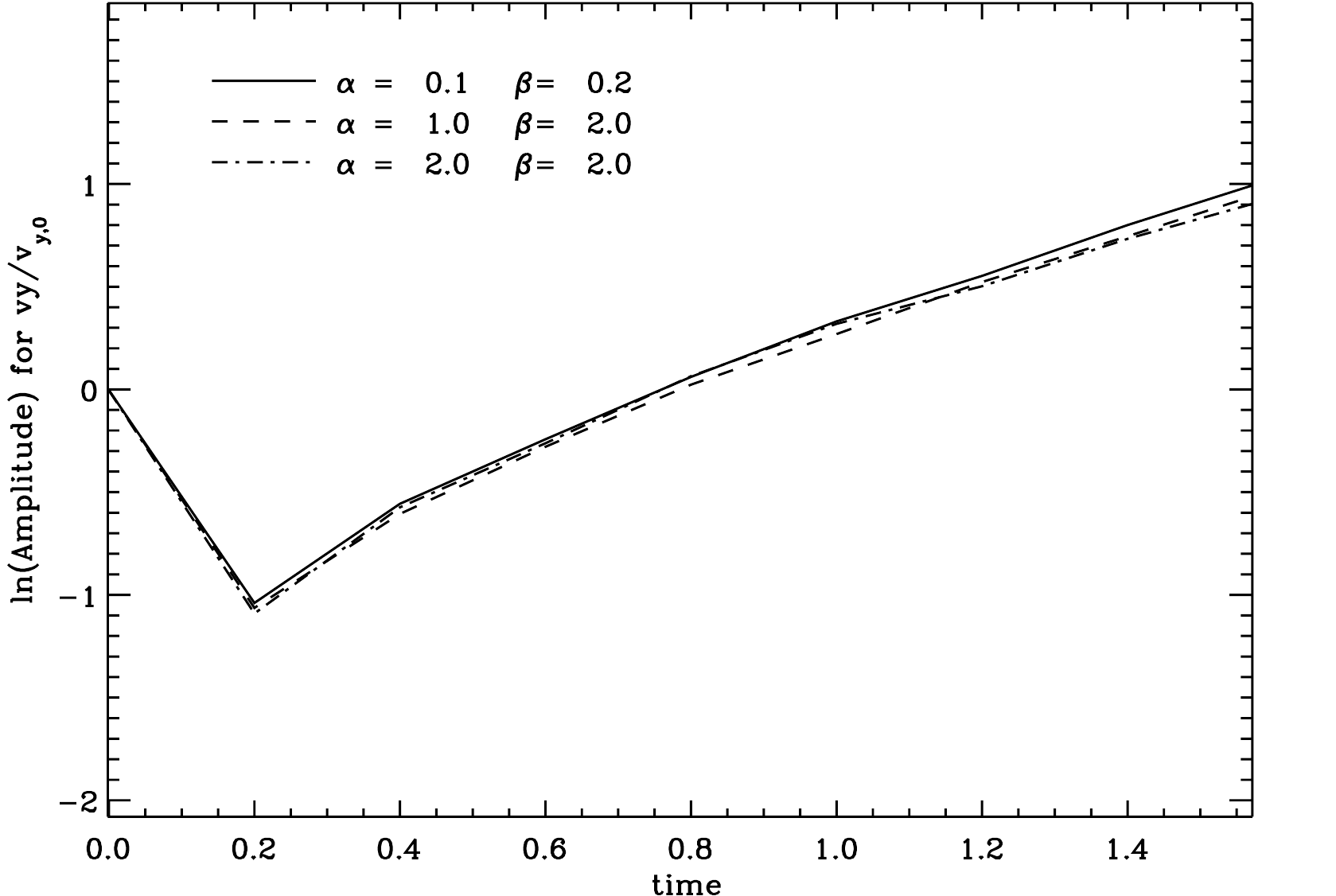}
\caption[]
{
  Time evolution of the VINE $v_y$-amplitude for 
  different values of the AV-parameters $\alpha$ and $\beta$, where
  the Balsara-viscosity has been used. The damping of the amplitudes is completely prohibited by the Balsara switch.
}
\label{vine_evolution_equ_dens_balsara}
\end{center}
\end{figure*}
\begin{itemize}
\item 
  {\sc{Dependence of KHI on $\alpha$, $\beta$:}} \\
  In Fig.~\ref{Shear_plots_alpha}, Fig.~\ref{Shear_plots_beta} and Fig.~\ref{multi_equal_alpha_beta_vy}
  we show the KHI-evolution for different values of
  $\alpha$ and $\beta$ without the Balsara-viscosity. 
  Increasing the AV-parameter $\alpha$ or $\beta$ results in a successive suppression of the KHI. 
  Values of $\alpha > 2$ and $\beta > 1$ lead to a decay of the
  initial perturbation. 
  However, $\beta$ does not affect the growth 
  as much as $\alpha$. Therefore, we first concentrate on
  $\alpha$ as the operating term on the KHI. \\
  Can we assign an equivalent physical viscosity $\nu_{\mathrm{SPH}}$
  to the SPH scheme, i.e. can we determine how "viscous" the fluid 
  described by SPH is intrinsically? 
  To quantify its value, the analytical slope
  (Eq.~\ref{slope_equal_densities}), with the viscosity being the free parameter, is fitted to the simulated 
  growing amplitudes.
  We show the best fits for $\alpha=0.125$ and $\alpha=2$ in the left
  panel of Fig.~\ref{multi_equal_alpha_beta_vy}, for which 
  we find the intrinsic viscosity of $\nu_{\mathrm{SPH}}=0.07$ and
  $\nu_{\mathrm{SPH}}=0.1$. 
  Here we assumed the time range of $[0.2,1]$, for which we determine
  the fits, to be well in the linear regime. \\
  In Fig.~\ref{alpha_fit_visc} we present the derived values of $\nu_{\mathrm{SPH}}$ as a function of
  $\alpha$. In summary, $\nu_{\mathrm{SPH}} $
  increases linearly with increasing plotted box size is from
  $[-0.5,0.5]$ in both directions, the resolution is $512^2$. 
  $\alpha$, and the corresponding
  slope is $0.039$. We also derive an offset of $0.065$, 
  which is the remaining intrinsic viscosity for $\alpha=0$. 
  For each simulation, we also show the effective Re number of the
  flow (see Fig.~\ref{alpha_fit_visc}, right y-axis), 
  which was computed from $Re = L \cdot U/\nu_{\mathrm{SPH}}$.
  The parameter $L$ describes the characteristic scale of the
  perturbation, in our case the wavelength and $U$ 
  is the velocity of the flow. 
  Clearly, the Reynolds-numbers we reach with our models are well below the commonly expected numbers
  for turbulent flows ($Re>10^5$). \\
  The effective viscosity of the flow is also influenced by different
  values of $\beta$. Changing $\beta$ by a factor of two (e.g. from
  $\beta=0.5$ to $\beta=1$) results in an
  increase in effective viscosity by a factor of $0.01$ (see right
  panel of Fig.~\ref{multi_equal_alpha_beta_vy}). 
\item 
  {\sc{Dependence on the Balsara-viscosity:}}\\
  We showed that AV leads to artificial viscous dissipation, resulting in the damping of the KHI. 
  To prevent this, we use the Balsara-viscosity, see also section~\ref{numdescVINE}. 
  In Fig.~\ref{vine_evolution_equ_dens_balsara} we show the corresponding
  amplitudes for three examples of 
  AVs: ($\alpha=0.1$, $\beta=0.2$), ($\alpha=1$, $\beta=2$) and 
  ($\alpha=2$, $\beta=2$). 
  Clearly, the Balsara viscosity reduces the damping of the KHI,
  rendering $\nu_{\mathrm{SPH}}$ almost independently of $\alpha$ and
  $\beta$ (see also Fig.~\ref{alpha_fit_visc}). 
\end{itemize}
\begin{figure*}
\begin{center}
\includegraphics[width=0.7\textwidth]{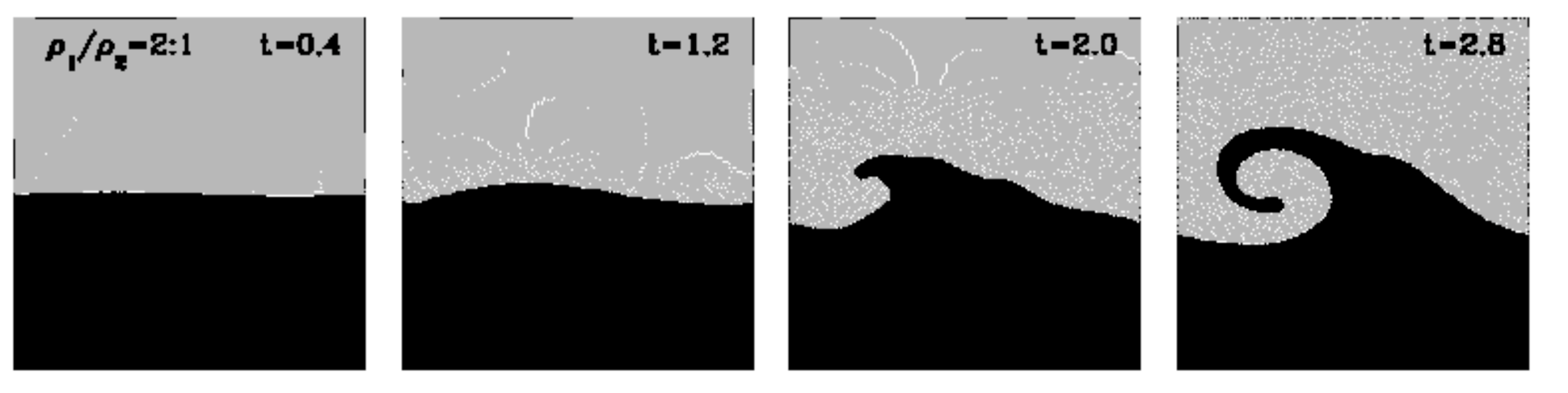}
\includegraphics[width=0.7\textwidth]{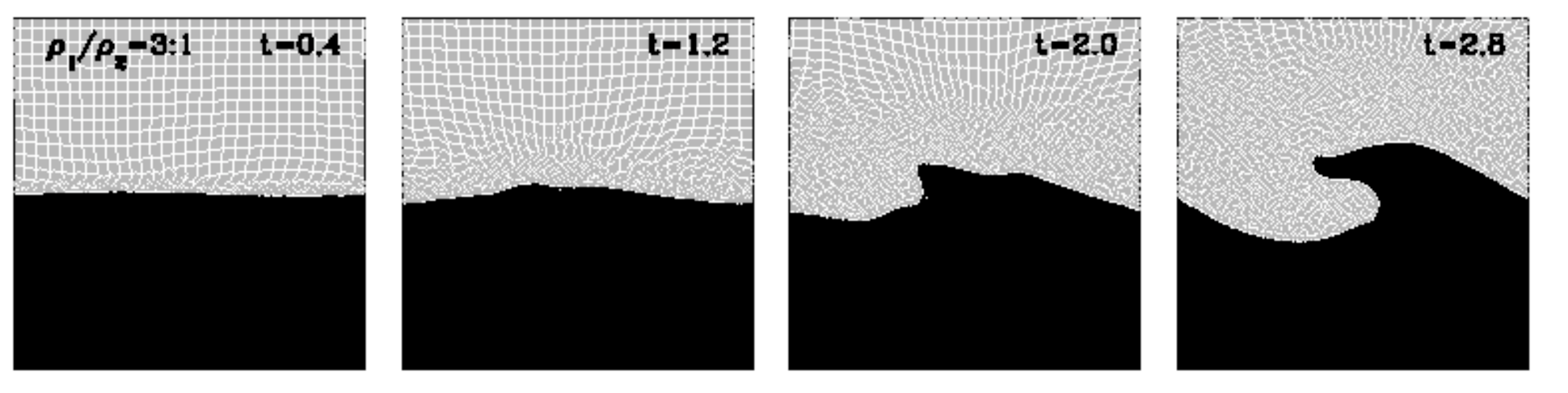}
\includegraphics[width=0.7\textwidth]{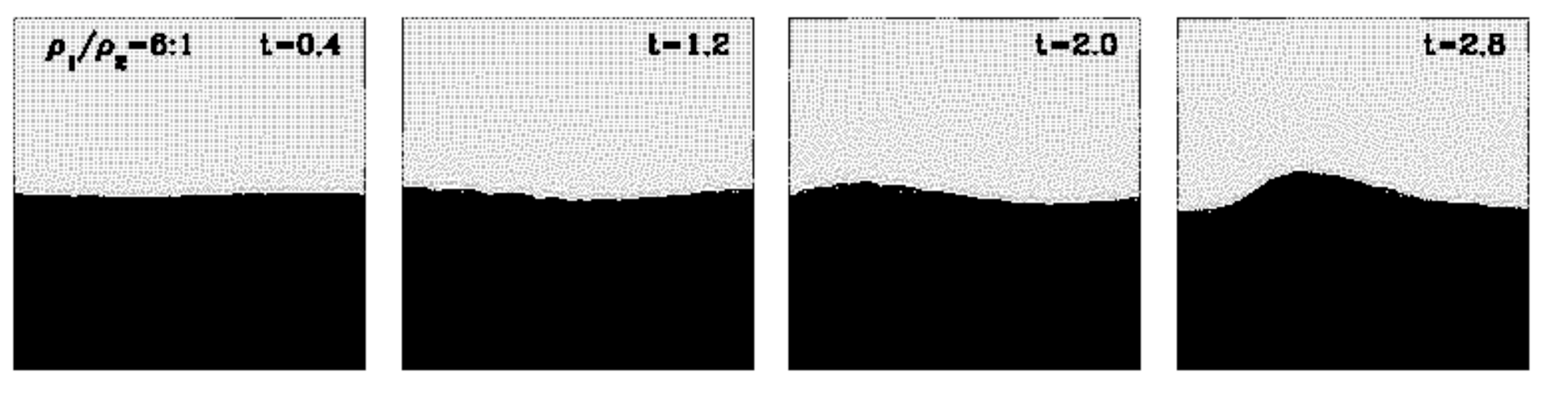}
\caption[Evolution and growth of the KHI for different density contrasts]
{
  Like Fig.~\ref{Shear_flow_diff} top panel, but for different density
  contrasts. From top to bottom we show DC=$2,\,3,\,6$. 
  For $DC\ge6$ the KHI does not develop anymore. 
}
\label{KHI_DC_evolution}
\end{center}
\end{figure*}
\begin{figure*}
\begin{center}
\includegraphics[width=0.49\textwidth]{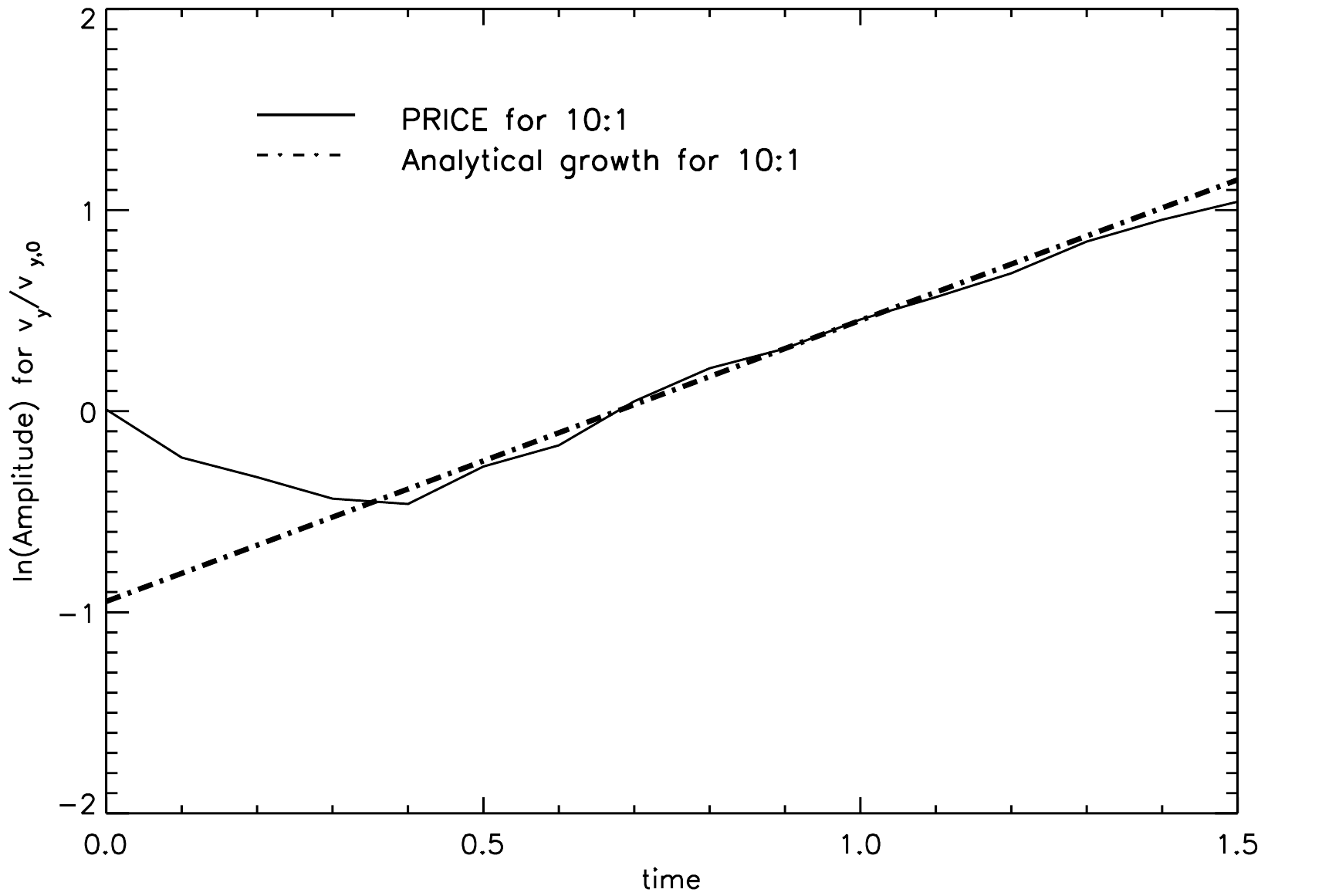}
\caption[]
{
  Time evolution of the KHI modeled with P08 for the $DC=10$. 
  The dashed-dotted line corresponds to the analytical prediction,
  Eq.~\ref{KHI_analytical_diff}, which is in good agreement with the simulation.
}
\label{PRICE_2to1_growth}
\end{center}
\end{figure*}
\begin{figure*}
\begin{center}
\includegraphics[width=0.7\textwidth]{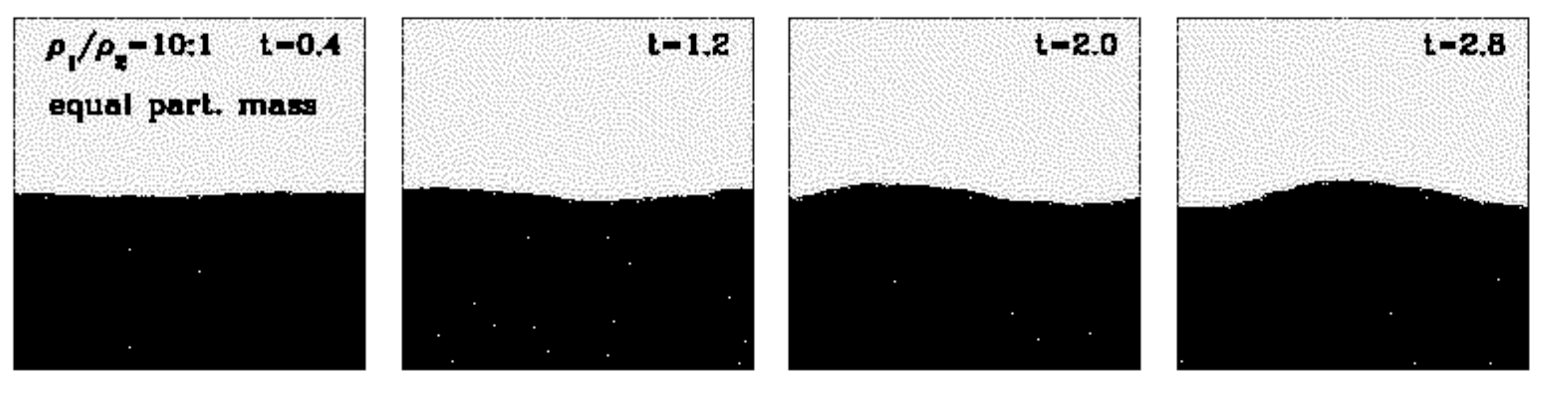}
\includegraphics[width=0.7\textwidth]{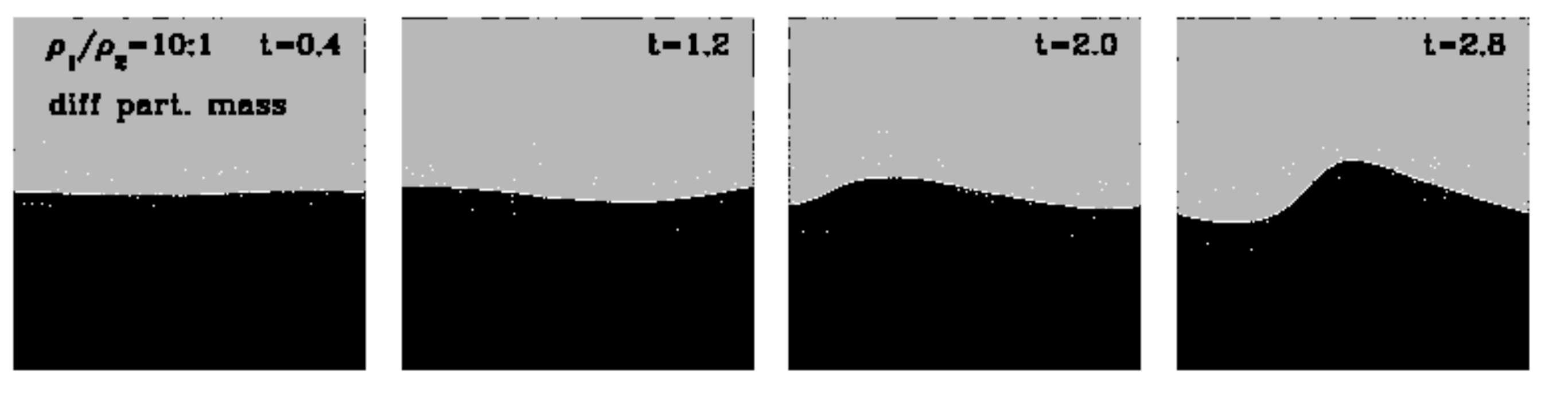}
\caption[Growth and evolution of KHI with variable density layers]
{
  KHI with VINE for $DC=10$. 
  Top: Case of equal particle masses.
  Bottom: Case of unequal particle masses and therefore equal particle
  numbers in both layers. The KHI is suppressed in all cases. 
}
\label{Shear_flow_diff}
\end{center}
\end{figure*}
\subsection{Fluid layers with variable densities:}
\label{SPH_diff_dens_layers}
While the previously addressed case of equal densities helped us to understand the
detailed evolution of the KHI as modeled with SPH, the astrophysically more interesting
case are shear flows with different densities. 
The resolution of the diffuse region is lower by a factor of
$\sqrt{DC}$, where $DC$ is the ratio of the densities in dense and
diffuse medium (e.g. $DC=10$ corresponds to a density contrast of $10:1$). 
We return to our standard set of parameters, in which case
$\alpha=0.1$ and $\beta=0.2$. 
For these low AV parameters we do not need the Balsara-viscosity (see
\ref{SPH_simulations_equal_density}). 
(Nonetheless, we did run test simulations with the Balsara switch,
which we found to confirm our 
former finding, since the growth of the KHI was not affected).
In the following, we (i) analyze the growth of the KHI for different
values of DC (with equal mass particles) and address the problem
of KHI suppression, while in (ii) we test the influence of equal mass or spatial
resolution. \\\\
(i) {{\sc KHI growth as a function of $DC$:}}\\
We show the KHI evolution for increasing $DC$ in
Fig.~\ref{KHI_DC_evolution}. 
For $DC\ge 6$ the KHI does not develop anymore. 
This SPH problem of KHI suppression has been 
studied in great detail (e.g. \citealp{Agertz_2007,Price_2007,Wadsley_2008,Read_Hayfield_Agertz_2009,Abel_2010}).
SPH particles located at the
interface have neighbors at both sides of the
boundary (i.e. from the dense- and less dense region). Therefore, the
density at the boundary is smoothed during the evolution. However, 
the corresponding entropy (or, depending on the specific code, the
thermal energy) is artificially fixed in these (isothermal) setups
which results in an artificial contribution to the SPH pressure force term, due to which the two
layers are driven apart. 
One possible solution is to either adjust the density (\citealp{Ritchie_2001,Read_Hayfield_Agertz_2009}), or to
smooth the entropy (thermal energy)
(\citealp{Price_2007,Wadsley_2008,Abel_2010}). \\
A remedy has been discussed by \cite{Price_2007}, who proposed to add 
a diffusion term, which is called artificial thermal conductivity (ATC), 
to adjust the thermal energy. (For a detailed study of ATC see \citealp{Price_2007}).
With this method, the KHI should develop according to the test cases of P08.\\
In Fig.~\ref{PRICE_2to1_growth} we test whether the P08 approach is
indeed in agreement with our analytical prediction. Note that P08 has a method
implemented to account for the artificial viscous dissipation
caused by AV (similar to the Balsara-viscosity). Thus, the viscous effects of AV are strongly reduced.
For $DC=10$ and using $512^2$ particles in the dense layer we indeed find good
agreement between measured and analytical growth rates. If
the standard SPH scheme is used, a correction term like ATC has to be
included to obtain a KHI in shear flows with different
densities, which is consistent with the analytical prediction.\\\\
(ii) {{\sc KHI growth using equal and different particle masses:}}\\
First, we investigate the development of the KHI for the standard SPH
case of equal mass resolution throughout 
the computational domain, and therefore fewer particles in the low
density fluid layer (see top panel of Fig.~\ref{Shear_flow_diff} for
$DC=10$, where the dense medium is resolved with $512^2$ particles). 
This results in a varying spatial resolution, due to the fact that 
SPH derives the hydrodynamic quantities within a smoothing length $h$ set by
a fixed number of nearest neighbors. This construct -- 
as has been discussed in detail earlier in e.g. Agertz et al. 2007 -- 
specifically lowers the Reynolds-number of the shear flow across density
discontinuities, thus affecting the evolution of the KHI. 
As can be seen in the top panel of Fig.~\ref{Shear_flow_diff}, the KHI
is completely suppressed. \\ 
Second, we test the case of equal spatial resolution in both fluid
layers, and therefore unequal particle masses within the 
computational domain (Fig.~\ref{Shear_flow_diff}, lower panel). 
Again, we find the KHI to grow too slowly with respect to the
analytical estimate. However, the suppression is less effectively in the latter case.
\section{GRID-Simulations of the KHI}
\label{Grid_simulations}
For comparison to the SPH treatment of Kelvin-Helmholtz instabilities,
we study an identical setup of fluid layers with the grid-based codes 
FLASH and PLUTO (see \S\ref{numdescFLASH}). 
We reuse the previously specified initial conditions with a grid
resolution of $512^2$ cells in the standard case. For FLASH, we
additionally include physical viscosity of various strenghth in some
of the simulations (see \S\ref{numdescFLASH}).
Note, that for the following examples we use $\sigma_0=1$ if not 
otherwise specified, which does not affect the growth of the amplitudes in the linear regime
(for further information see discussion in the
Appendix~\ref{dependence_sigma_0}). 
\subsection{Fluid layers with equal densities}
\label{FLASH_equal}
\subsubsection{Non-viscous evolution}
\label{nonviscous_equal_evolution}
\begin{figure*}
\begin{center}
\includegraphics[width=0.49\textwidth]{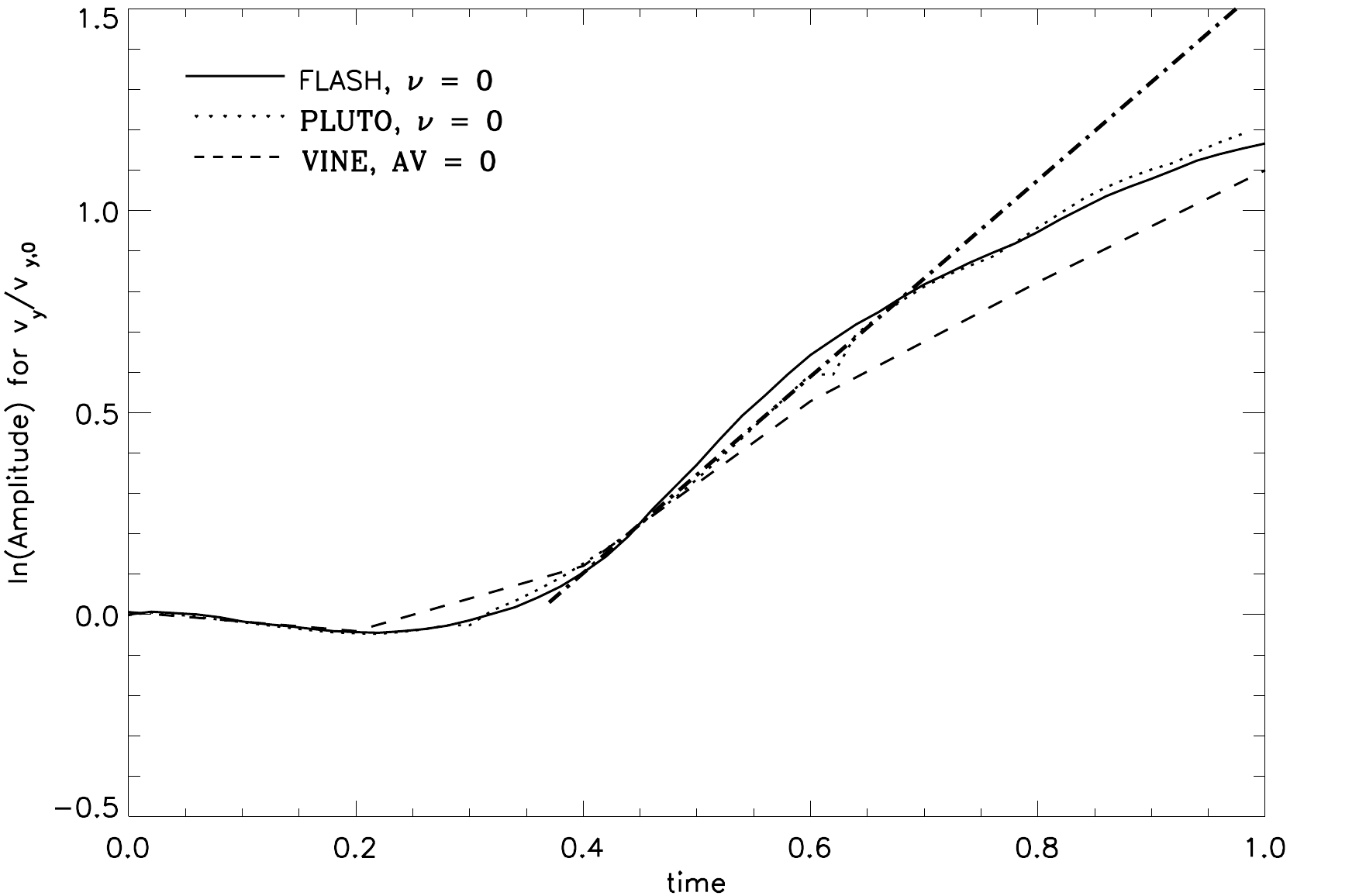}
\includegraphics[width=0.49\textwidth]{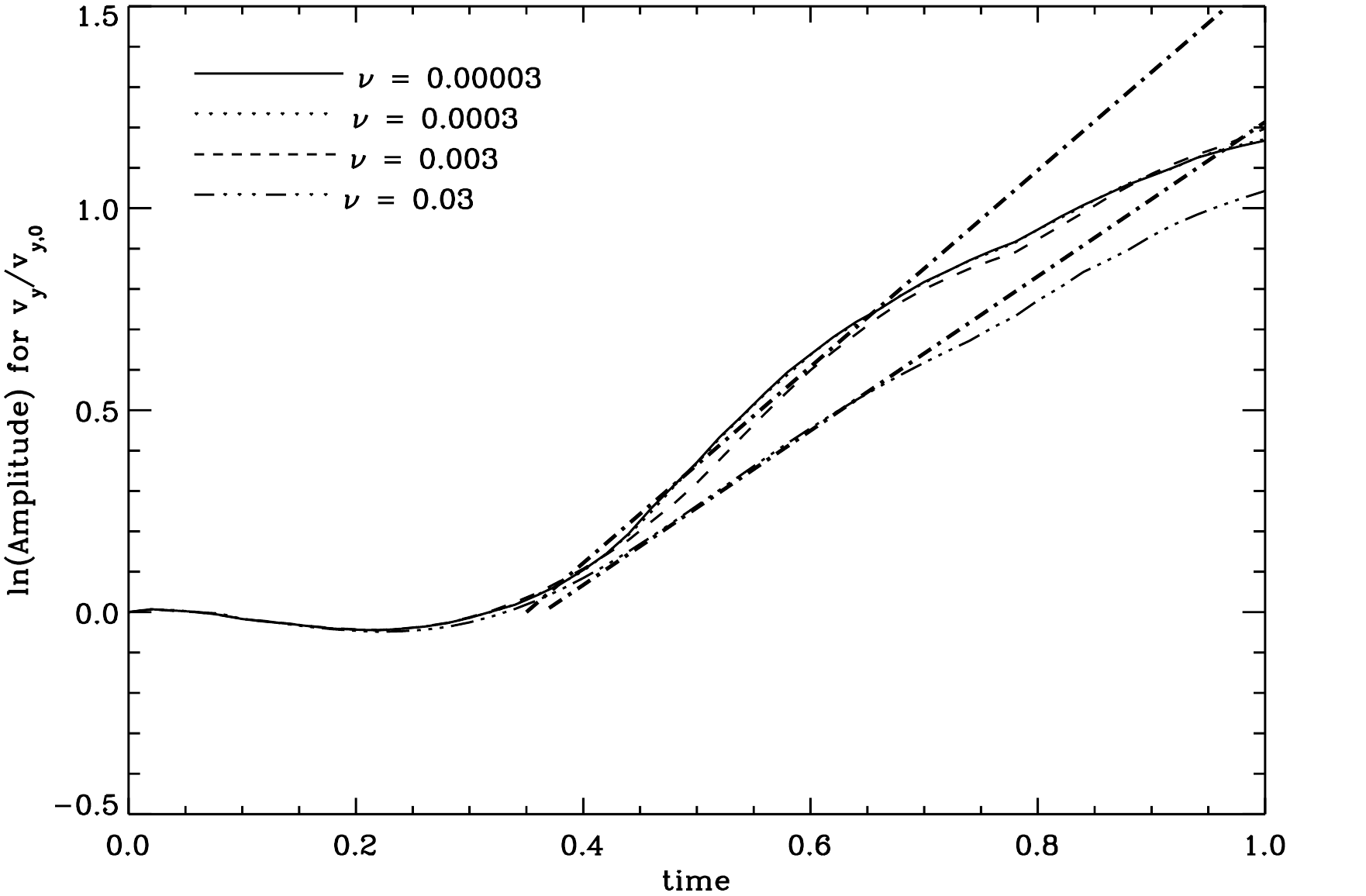}
\caption[]
{
  Evolution of KHI amplitudes for equal density
  layers. Left panel: Non-viscous evolution for FLASH
  (solid line), and PLUTO (dotted line). Additionally, we show the example
  with VINE (dashed line), where the AV has been set to zero
  ($\alpha=\beta=0$).
  Right panel: Viscous KHI evolution using FLASH. 
  The thick dashed-dotted lines correspond to the analytical
  prediction, Eq.~\ref{slope_equal_densities}. 
}
\label{equal_dens_novisc_flash_pluto}
\end{center}
\end{figure*}
The left panel of Fig.~\ref{equal_dens_novisc_flash_pluto} shows the 
non-viscous KHI-evolution, using FLASH (solid line), PLUTO (dotted
line), and for comparison VINE (dashed line). In the VINE
example, the AV has been set to zero ($\alpha=\beta=0$). 
The expected analytical 
growth (Eq.~\ref{slope_equal_densities}) reduces with $\nu=0$ to $n\sim k \cdot U =
2.43$ (indicated by the thick dashed dotted line).
The FLASH and PLUTO amplitudes develop in a similar pattern and are
almost undistinguishable. 
Their fitted slopes within the linear regime (which lies roughly between $t=0.3 - 0.6$) results in $n_{\mathrm{fit}}=2.49$. 
FLASH and PLUTO show a consistent growth in agreement with the analytical prediction. 
VINE on the other hand exhibits a slightly slower growth. This
deviation is due to the intrinsic viscosity ($\nu_{\mathrm{int}}=0.065$) that was estimated in \ref{SPH_simulations_equal_density}. 
\subsubsection{Viscous evolution}
\label{viscous_equal_evolution}
\begin{figure*}
\begin{center}
\includegraphics[width=0.7\textwidth]{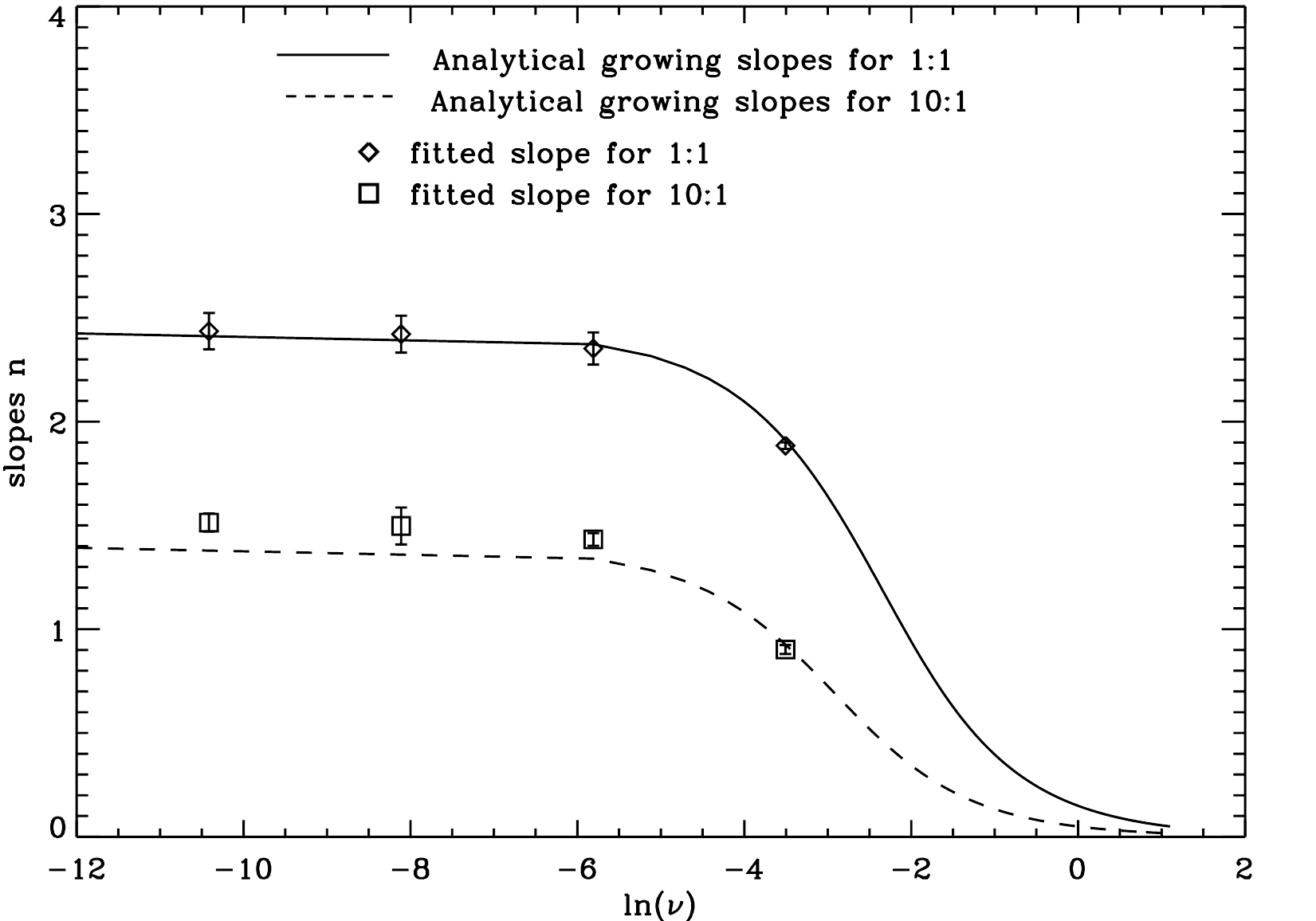}
\caption[]
{
  Comparison of the analytical expectation and the models for DC=$1$
  (diamond shaped symbols) and $DC=10$ (square symbols). 
  The slopes derived for FLASH correspond to the analytical fits.
  The lines represent the analytic prediction, 
  for DC=$1$ (solid line, see Eq.~\ref{slope_equal_densities}) 
  and $DC=10$ (dashed line, see Eq.~\ref{mode_diff_layers}).
}
\label{Slopes_Flash}
\end{center}
\end{figure*}

The right panel of Fig.~\ref{equal_dens_novisc_flash_pluto} shows the 
viscous KHI-amplitudes using FLASH. 
The corresponding analytical predictions (Eq.~\ref{slope_equal_densities}) are shown by
the thick dashed-dotted lines for the examples with $\nu=0.00003$ and $\nu=0.03$. 
To quantify the growth of the KHI in the FLASH simulations, 
we again fit the slopes of the KHI-amplitude in the linear regime
(between $t=0.3 - 0.6$). The result along with the corresponding error is plotted in
Fig.~\ref{Slopes_Flash}. 
For small viscosities ($\nu<0.003$), we find the growth rates of the KHI in FLASH to
be in good agreement with the analytical prediction. 
In this viscosity range, the dominant term in the analytical prediction (Eq.~\ref{slope_equal_densities}) is $\sim k U$. 
Therefore, any influence of $\nu$ is marginal, and the amplitudes do
not change considerably. FLASH treats the fluid as if $\nu \approx 0$. \\
However, with increasing viscosity, the amplitudes should be damped. 
This behavior is in fact visible in the right panel of
Fig.~\ref{equal_dens_novisc_flash_pluto} 
(as well as in Fig.~\ref{Slopes_Flash}). The growth rates of the KHI
agree very well with the analytical prediction.
\begin{figure*}
\begin{center}
\includegraphics[width=1.0\textwidth]{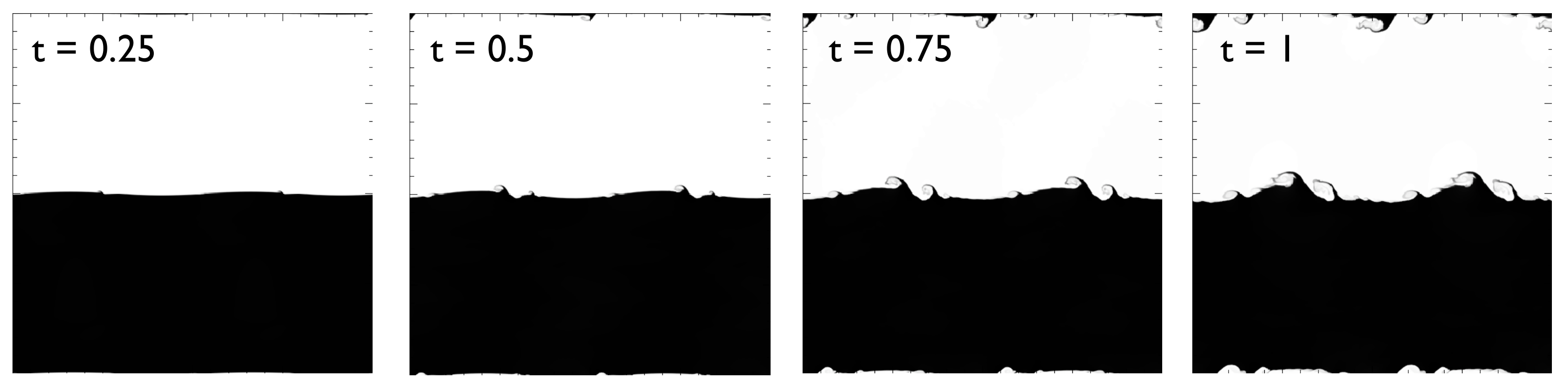}
\includegraphics[width=1.0\textwidth]{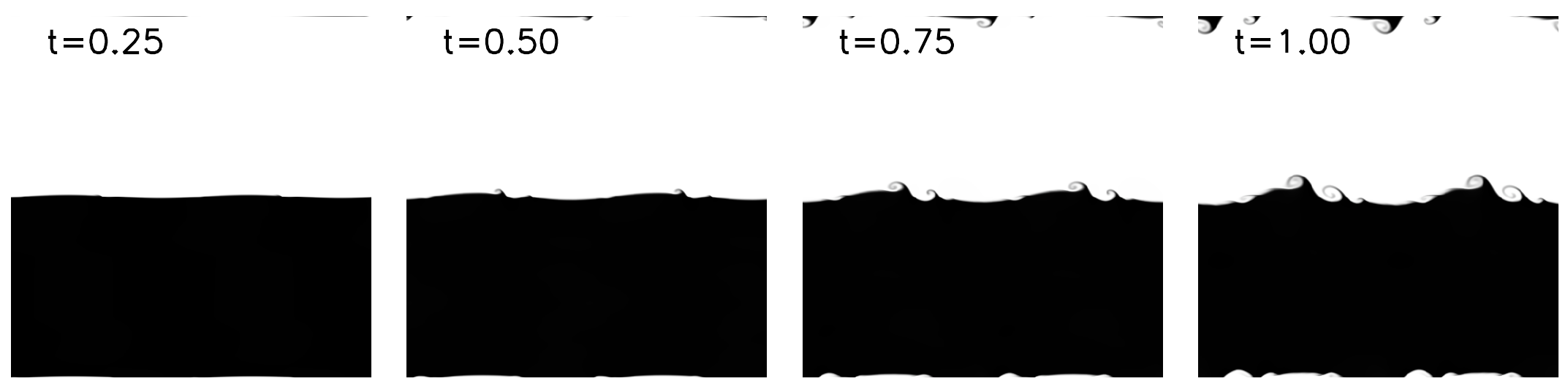}
\caption[]
{
  Time evolution of the KHI density in a simulation with nu=0 and
  $DC=10$ for FLASH (top row) and PLUTO (bottom row). The plotted box size is from $[-1,1]$ in both directions, the 
  resolution is $512^2$. The KHI develops, which is in contrast to the
  example simulated with VINE.
}
\label{FLASH_diff_dens_evolution}
\end{center}
\end{figure*}
\begin{figure*}
\begin{center}
\includegraphics[width=0.49\textwidth]{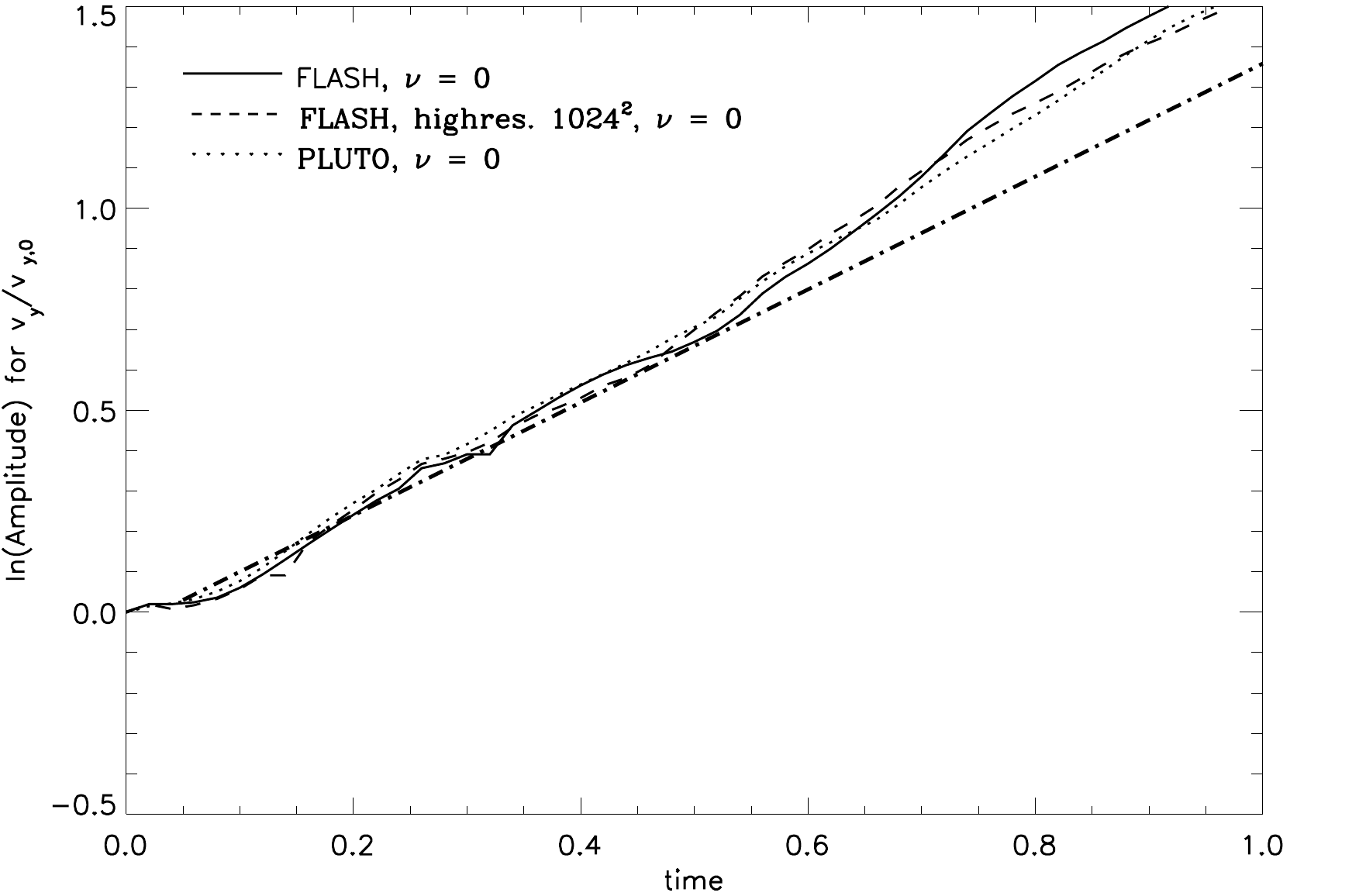}
\includegraphics[width=0.49\textwidth]{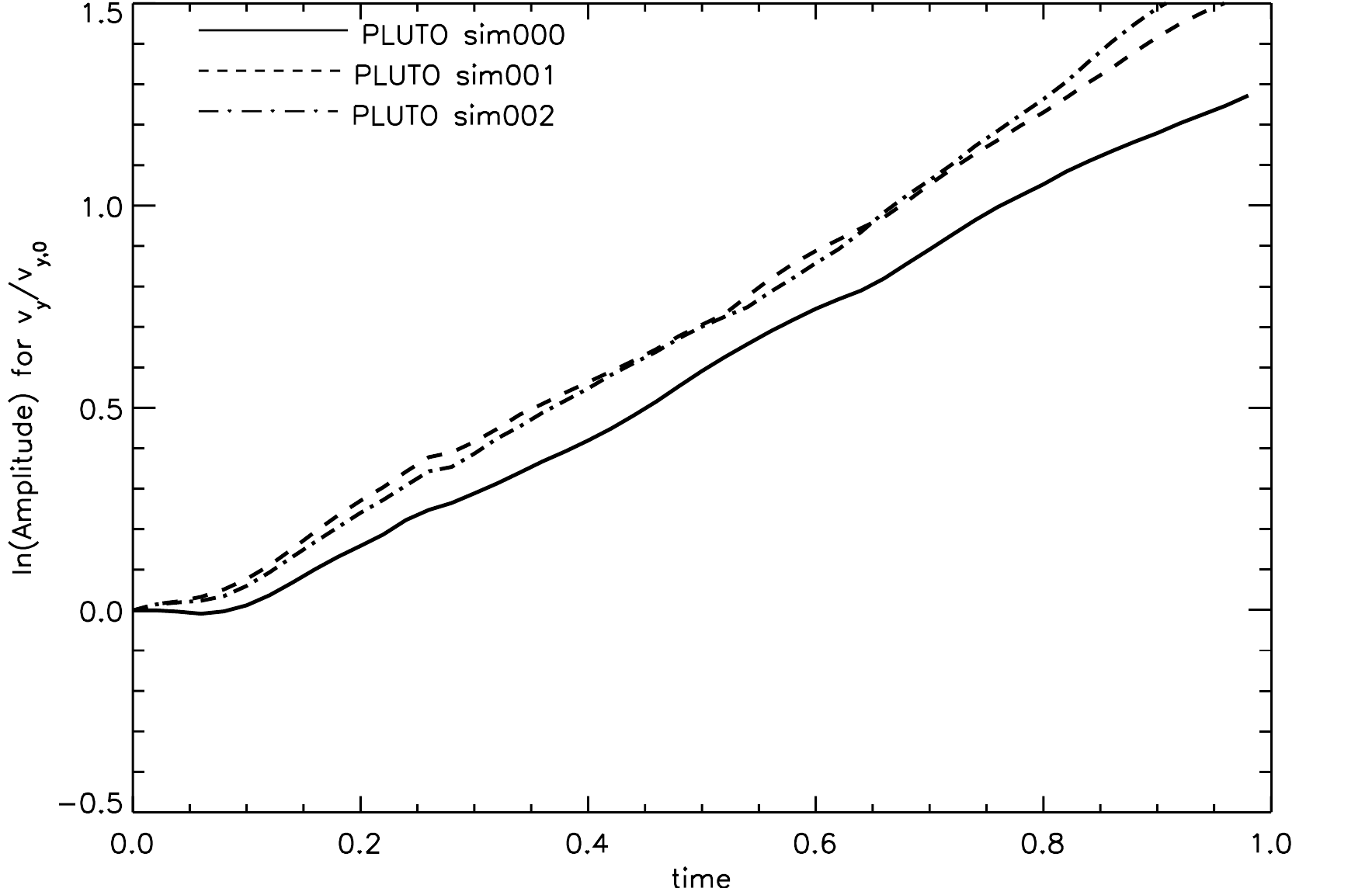}
\caption[]
{
  The same as in Fig.~\ref{equal_dens_novisc_flash_pluto} but for a $DC=10$. 
  Left panel: Non-viscous evolution for FLASH (solid line), PLUTO
  (dotted line) and the high-resolution ($1024^2$) amplitude for FLASH
  (dashed line). 
  Right panel: Non-viscous evolution using PLUTO, with different solvers, see text for more details.
}
\label{viscous_evolution_flash}
\end{center}
\end{figure*}
\begin{figure*}
\begin{center}
\includegraphics[width=0.49\textwidth]{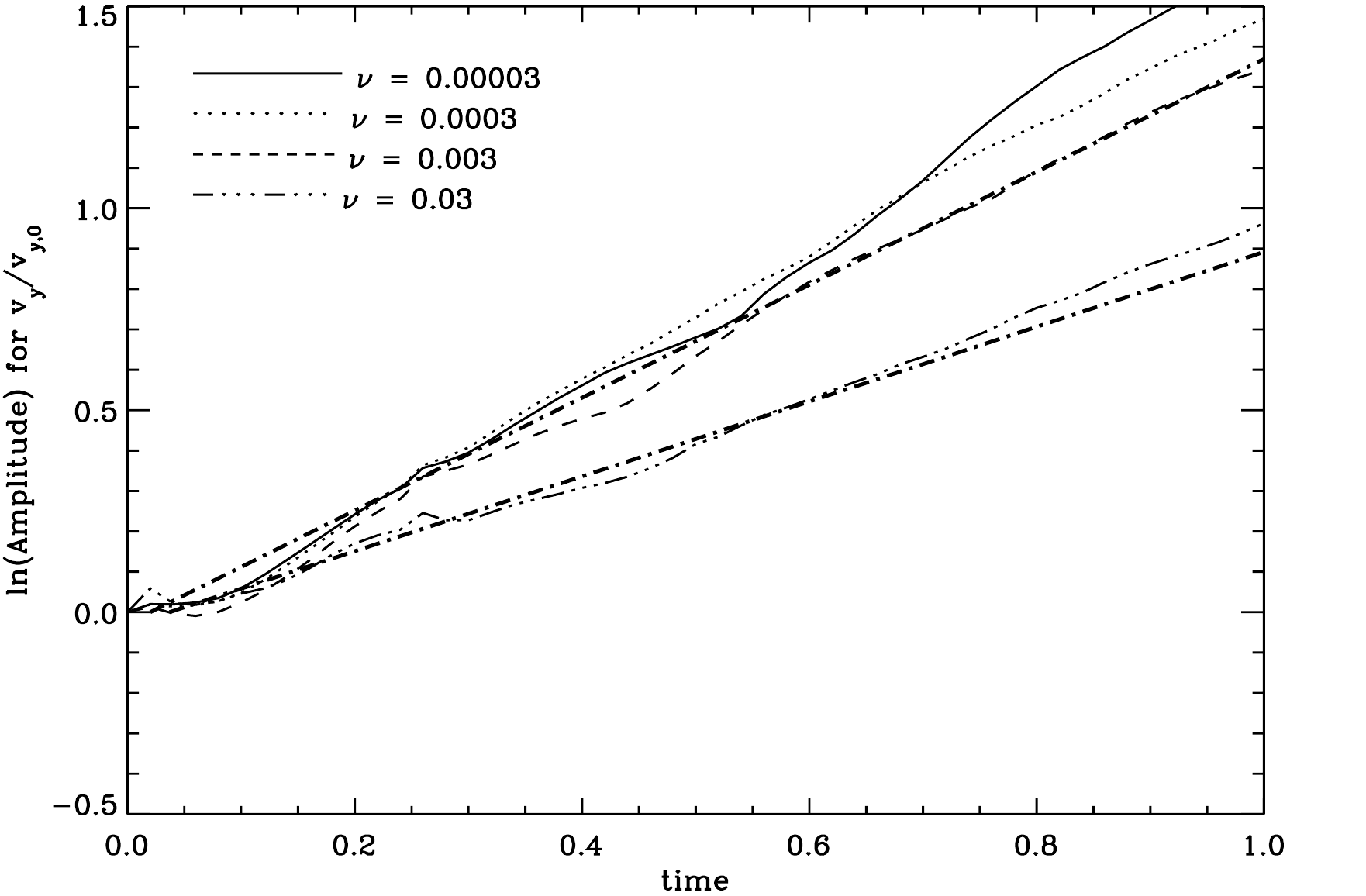}
\caption[]
{
  Viscous evolution using FLASH. The thick dashed-dotted lines correspond to the analytical 
  prediction, Eq.~\ref{mode_diff_layers}.
}
\label{viscous_evolution_flash_new}
\end{center}
\end{figure*}
\subsection{Fluid layers with different densities}
\label{FLASH_diff}
\subsubsection{Non-viscous evolution}
\label{nonviscous_diff_evolution}
Finally, we investigate a density contrast of $10 : 1$, similar to the
example studied with VINE (see \S~\ref{SPH_diff_dens_layers}). 
Fig.~\ref{FLASH_diff_dens_evolution} shows the non-viscous evolution of the KHI
for the $DC =10$ case (upper line for FLASH, bottom line for PLUTO). 
It can be seen that for both codes the interface layer starts to roll-up and the instability 
is developed. This is in disagreement with the previously discussed case using SPH, where the KHI is completely 
suppressed for DC$>6$ (see \ref{SPH_diff_dens_layers}).\\
The left panel of Fig.~\ref{viscous_evolution_flash} presents the corresponding
amplitudes for FLASH (solid line) and PLUTO (dotted line) compared to the analytical
prediction (thick dashed-dotted line),  which in this case reduces to
\begin{equation}
\label{KHI_analytical_diff}
n = \pm i\sqrt{4k^2U^2\alpha_1\alpha_2}.
\end{equation}
For FLASH we show two different resolutions ($512^2$ and $1024^2$). 
The amplitudes resulting in the case of low and high resolution are effectively indistinguishable. 
This is an important result, as it demonstrates that small scale
perturbations, which arise due to numerical noise and which could
violate the linear analysis (as we then might follow the 
growth of higher order modes rather than the initial perturbation) are
not important. Therefore, we have shown that our simulations are
converged as we would otherwise expect the growth of the KHI to be slightly dependent on the
grid resolution (see e.g. the recent findings of \cite{Robertson_2009}, who had to
smooth the density gradient between the two fluid layers in order 
to achieve convergence in terms of grid resolution).
Moreover, both FLASH and PLUTO evolve similarly. For all three
examples the slope of the amplitude evolution can be approximated to
$1.4$, which is in good agreement with the analytical expectation.
Note that we do not show the comparison with the VINE amplitude since the KHI
does not evolve for $DC=10$ (see \ref{SPH_diff_dens_layers}). \\
Many grid codes offer a variety of hydrodynamical solvers. We
therefore tested the influence of different numerical schemes on
the growth of the KHI using PLUTO (see right panel of
Fig.~\ref{viscous_evolution_flash} ). 
We show three different examples; 'sim000' is a Lax-Friedrichs scheme 
together with a second order Runge-Kutta solver (tvdlf); 'sim001'
implements a two-shock Riemann solver with linear reconstruction
embedded in a second order Runge-Kutta scheme; 
'sim002' also implements a two-shock Riemann solver, but with
parabolic reconstruction, and embedded in a third order Runge-Kutta
scheme. Both, 'sim001' and 'sim002' show a similar growth of the KHI
in agreement with the analytical prediction (see
Fig.~\ref{viscous_evolution_flash}, top right panel). The more
diffusive scheme used in 'sim000' causes a small delay in the growth
of the KHI, but results in a similar slope within the linear regime (up to $t=0.6$).
\subsubsection{Viscous evolution}
\label{Viscous_diff_evolution}
Fig.~\ref{viscous_evolution_flash_new} shows the 
viscous KHI-amplitudes using FLASH, which are increasingly suppressed with $\nu$. 
The corresponding analytical prediction (Eq.~\ref{mode_diff_layers})
is shown for $\nu=0.0003$, and $\nu=0.03$ (thick dashed-dotted lines). 
For $\nu < 0.03$ the simulated growth rate is slightly enhanced by a factor of $\sim 0.12$ as compared to the
analytical prediction (see also Fig.~\ref{Slopes_Flash}). However, for
higher viscosities ($\nu \ge 0.03$) we find good agreement between simulation and analytical prediction.
\section{Conclusions}
\label{conclusions}
We have studied the Kelvin-Helmholtz instability applying different numerical schemes. 
We use two methods for our SPH models, namely the 
Tree-SPH code VINE \citep{Wetzstein_Nelson_Naab_Burkert_2008, Nelson_Wetzstein_Naab_2008}, 
and the code developed by \cite{Price_2007}.
The grid based simulations of the KHI rely on FLASH
\citep{Fryxell_2000}, while as a test for the non-viscous evolution we
also apply PLUTO \citep{Mignone_2007}. \\
We first extended the analytical prescription of the KHI by 
\cite{Chandrasekhar} to include a constant viscosity. With this
improvement we were able to measure the intrinsic viscosity of our
subsequently performed numerical simulations. We test both SPH as well
as grid codes with this method. \\
We then concentrated on the KHI-evolution with SPH. 
We performed a resolution study to measure the dependence of
the KHI growth on the mean number of SPH neighbors ($\bar{n}_{\mathrm SPH}$) and the total
number of particles, respectively. We found that our simulations were
well resolved and that a different number of $\bar{n}_{\mathrm SPH}$
did not significantly influence the KHI growth rate. \\
In case of equal density shearing layers we then measured the
intrinsic viscosity in VINE by evaluating our simulations against the analytical prediction in the linear regime.
Without using the Balsara viscosity the AV parameters $\alpha$ and $\beta$ effectively lead to a damping of the KHI. 
The commonly suggested and used settings of $\alpha=1$, and $\beta=2$
result in a strong suppression of the KHI. 
More quantitatively, we derive values of $0.065 < \nu_{\mathrm{SPH}} <
0.1$ for $0< \alpha < 1$. 
Different values of $\beta$ do not have a strong impact on
$\nu_{\mathrm{SPH}}$. 
By introducing the Balsara-viscosity the dissipative effects of the AV
can be reduced significantly, effectively rendering the results to be 
independent of $\alpha$ and $\beta$. However, the constant floor viscosity of $\nu_{\mathrm{SPH}}=0.065$ prevails.
Furthermore for a given $\alpha$, we estimated the effective
Reynolds-number ($Re$) of the flow. 
For the minimum SPH viscosity of $nu_{\mathrm{SPH}}=0.065$ 
we derive a maximum Reynolds number of $12$.
This is very small compared to typical Reynolds numbers of real turbulent flows ($Re > 10^5$). 
For different density shearing layers we confirmed the results
discussed in \cite{Agertz_2007}, i.e. the KHI is completely 
suppressed for shear flows with different densities (in the case of
VINE for $DC\ge 6$). 
Here, using the Balsara switch does not solve the problem. This
indicates that other changes to the SPH formalism are required in
order to correctly model shearing layers of different densities. To demonstrate this we applied the
solution of \cite{Price_2007} 
to our initial conditions for DC=10. In this case the KHI was
suppressed in VINE. However, we found good agreement between the
analytically predicted amplitude evolution and the simulation of
\cite{Price_2007} for $DC=10$. \\
The second part of this paper addresses the non-viscous- and viscous KHI
evolution using grid codes. 
In the case of equal density shearing layers, we found the non-viscous growth rates
for shear flows with FLASH and PLUTO to be in good agreement with the analytical prediction.
In the viscous case, the FLASH-amplitudes show only a minor dependency on the
viscosity if $\nu < 0.03$. 
Increasing the viscosity leads to a damped evolution, with the simulated
growth coinciding with the analytical prediction. \\
For non-viscous shear flows (with a density contrast of $DC=10$) the
KHI does develop for FLASH and PLUTO in agreement with the analytical
prediction. In the viscous case FLASH (also analyzed with $DC=10$)
slightly overpredicts the
corresponding growth rates for $\nu < 0.03$ by a constant factor of
$\sim 0.12$. \\
The comparison between VINE, FLASH and PLUTO in the equal density case, where $AV=0$ and $\nu=0$, demonstrated that VINE does 
have an intrinsic viscosity (which we estimated to $\nu_{\mathrm{int}}
\sim 0.065$). 
\section*{Acknowledgments}
I would like to thank Volker 
Springel for his useful suggestions, and Thorsten Naab for his support. Many thanks also to Oscar Agertz
for helpful discussions, as well as Eva Ntormousi.  
This research was supported by the DFG priority program
SPP 1177 and by the DFG cluster of excellence 'Origin and Structure of the Universe'. 
Part of the simulations were run on the local SGI ALTIX 3700 Bx2 which was also partly 
funded by this cluster of excellence.
FLASH was developed by the DOE-supported
ASC/Alliance Center for Astrophysical Thermonuclear Flashes at the
University of Chicago. S.Walch gratefully acknowledges the support of
the EC-funded Marie Curie Research Training Network {\sc
  Constellation} (MRTN-CT-2006-035890). M. Wetzstein gratefully
acknowledges support from NSF grant 0707731.   
\bibliographystyle{mn2e}
\bibliography{literature}
\appendix
\section{Dependence of KHI-amplitudes on $\sigma_0$}
\label{dependence_sigma_0}
\begin{figure*}
\begin{center}
\includegraphics[width=0.49\textwidth]{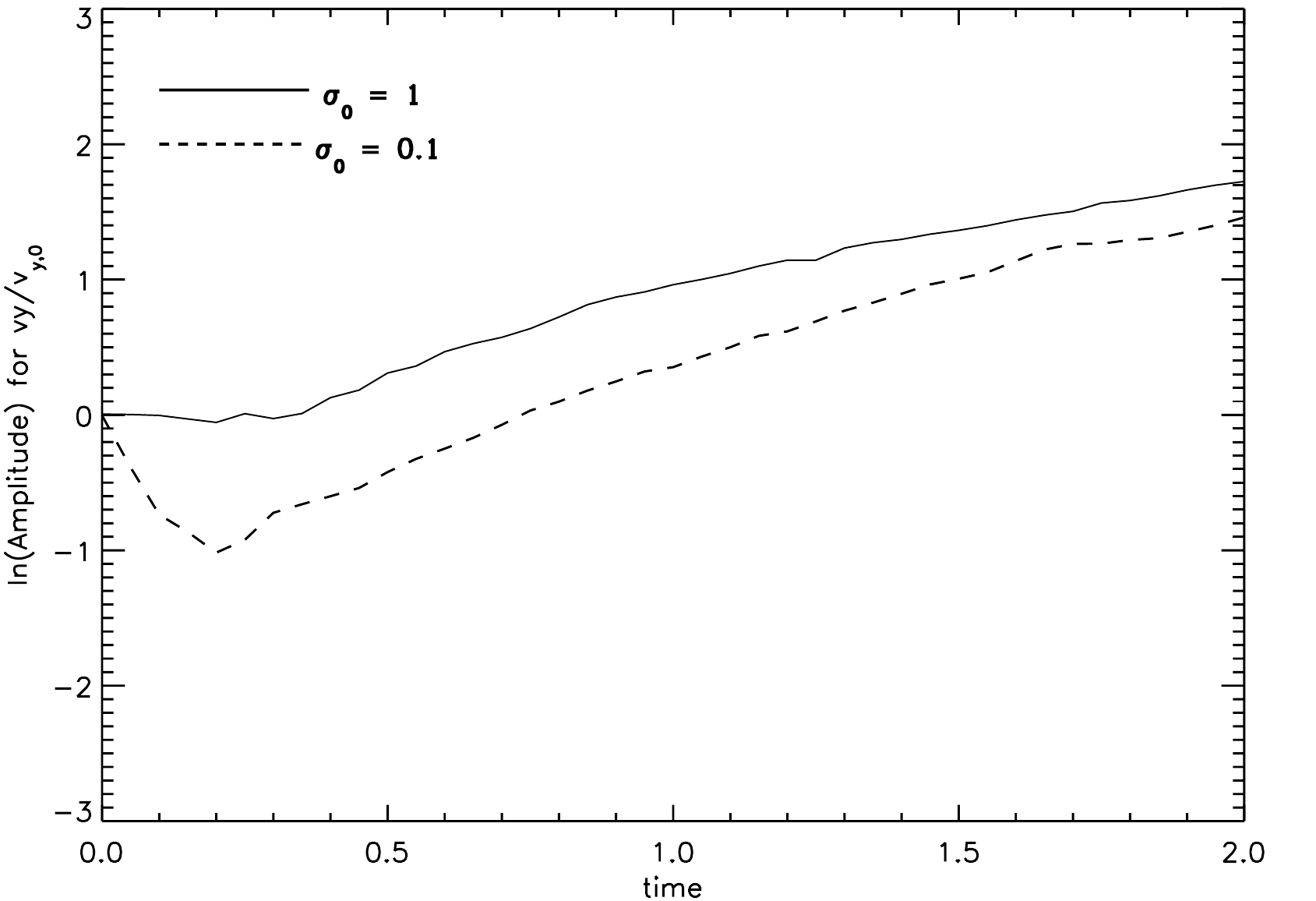}
\includegraphics[width=0.49\textwidth]{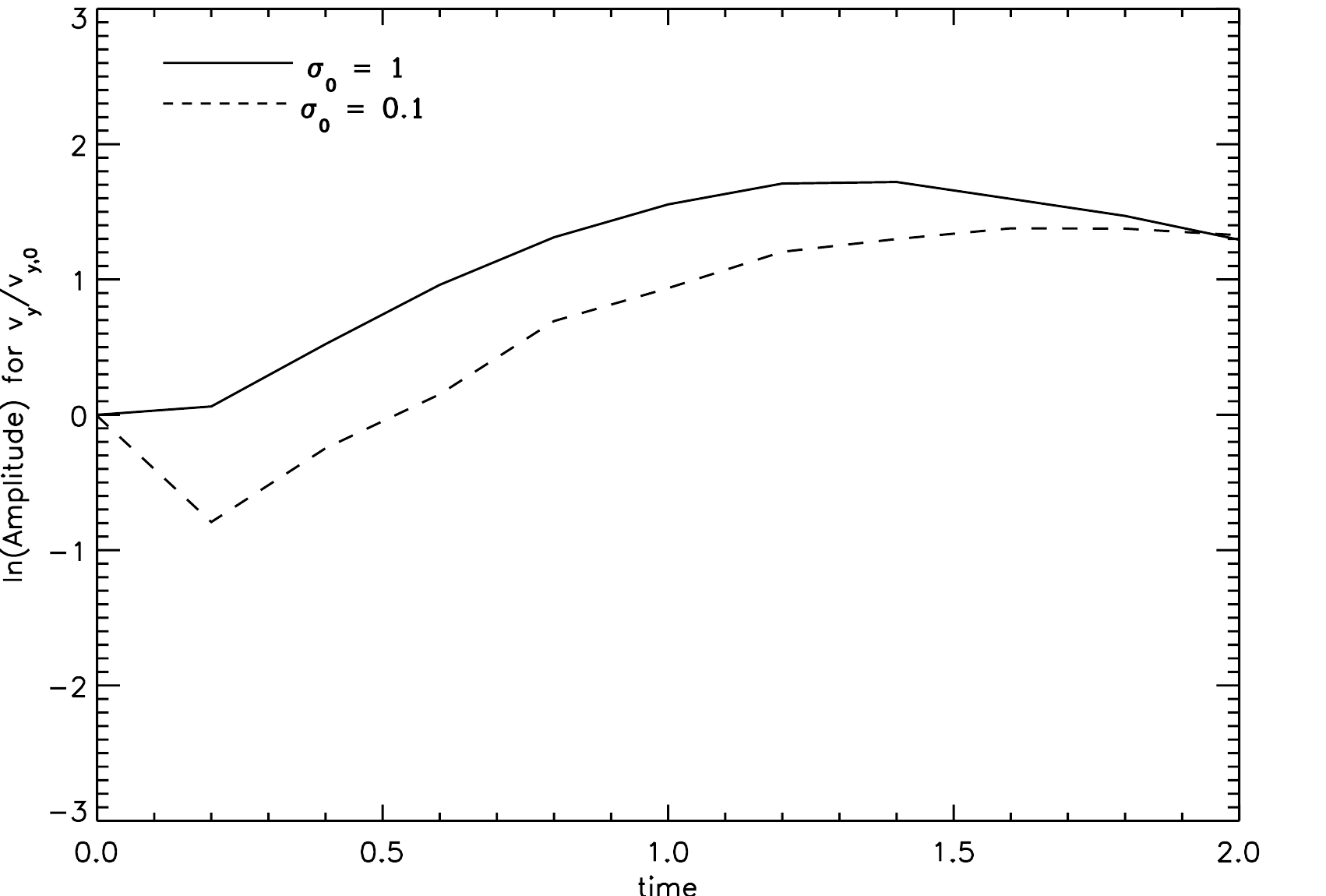}
\caption[Variation of KHI-amplitude for different values of $\sigma_{0}$]
{
Variation of KHI-amplitude in the case of equal density layers using VINE (left side) and FLASH (right side) for different values of $\sigma_{0}$. 
For Flash the viscosity has been set to $\nu=0.3$
}
\label{diff_sigma_0}
\end{center}
\end{figure*}
  {\sc{Dependence of KHI on $\sigma_0$:}} \\
  This parameter determines the strength of the initial
  $v_y$-perturbation (Eq.~\ref{velocity_perturbation}). 
  In Fig.~\ref{multi_equal_neighbors_vy} 
  we show the time evolution of the vy-amplitude, which describes the
  growth of the KHI. For $t \le 0.2$ the amplitudes decrease since the
  SPH particles lose kinetic energy by moving along the y-direction into the area of the opposite stream.
  If the magnitude of the initial perturbation is low (i.e. small
  $\sigma_0$), then the decrease in the amplitude is stronger than for
  e.g. $\sigma_0 = 1$, where the initial perturbation is large and
  the decrease less prominent. But independently of the value
  of $\sigma_0$ the subsequent growth of the instability is similar, and we obtain  
  comparable results neglecting the decreasing initial part. 
  Fig.~\ref{diff_sigma_0} shows the dependency of the KHI-amplitudes using different values of $\sigma_0$, for VINE (left side) and  
  FLASH (right side). For this example we use equal density layers, where for FLASH a viscosity of $\nu=0.3$ has been taken. 
  Clearly visible is the initial drop caused by a low value of $\sigma_0$. This is the case for both codes, and arises 
  due to the transformation of energy to build up the KHI. The 
  fitted slopes do not vary much with $\sigma_0$. To extract the slopes,
  we concentrate on the time evolution after this initial drop.
\section{Measuring the KHI-amplitudes}
\label{FT_KHI_amplitudes}
\begin{figure*}
\begin{center}
\includegraphics[width=0.7\textwidth]{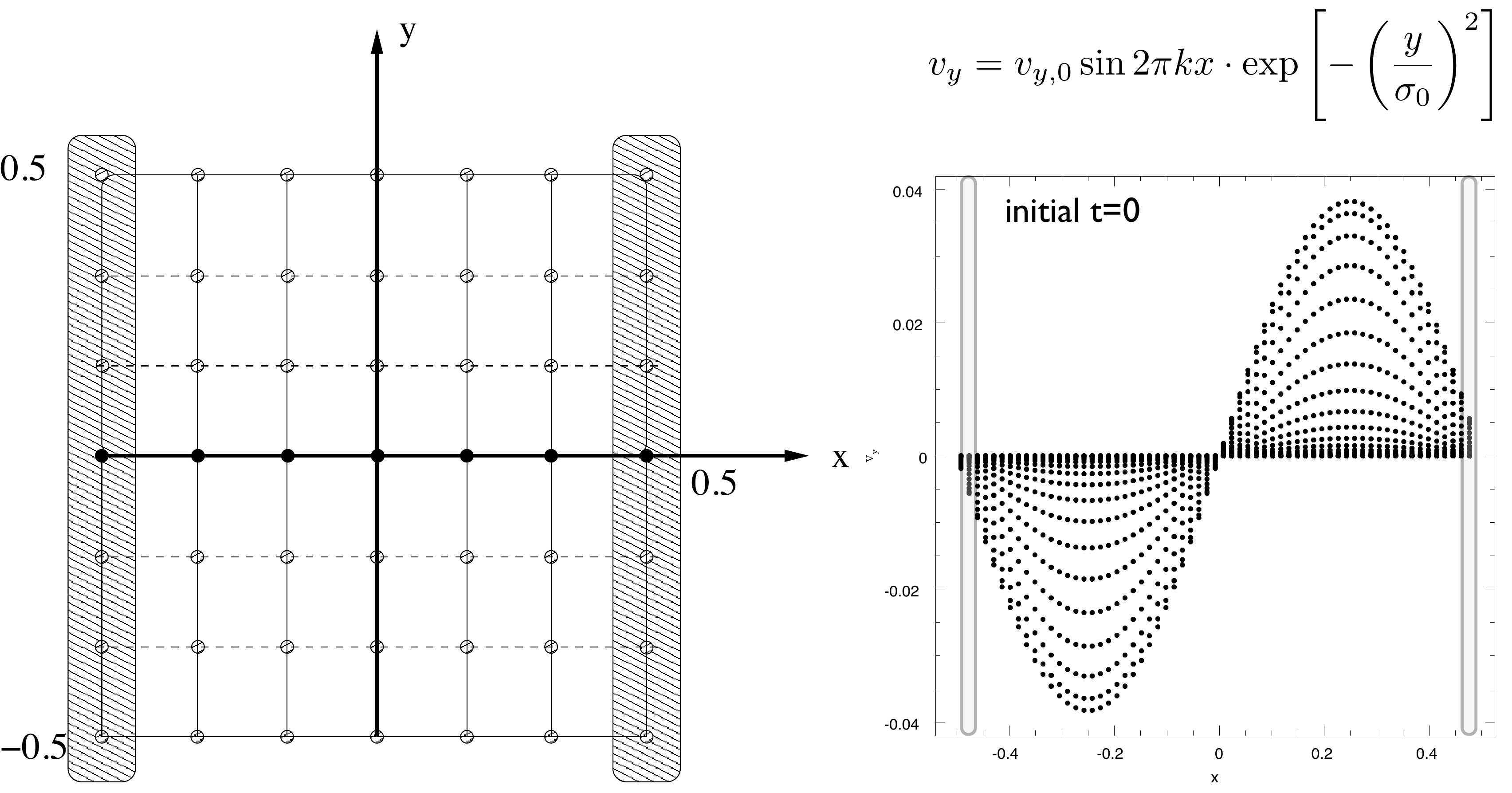}
\caption[FFT-Transformation of grid points]
{Method to measure the KHI-amplitudes: The $v_y$-velocity of the particles within the shaded region are 
subject to the Fourier-Transformation. The maximum of the Back-Transformation gives the maximal 
amplitude. }
\label{FFT_grid_points}
\end{center}
\end{figure*}
To measure the amplitude growth of the KHI, we apply a Fourier-Transformation (FT) to 
the $v_y$-velocity component of the grid points. The FT allows to select the desired modes 
reducing the numerical noise.\\
The region of our focus, $x=[-0.5, 0.5]$ and $y=[-0.5, 0.5]$ contains one 
mode of the $v_y$-perturbation (Eq.~\ref{velocity_perturbation}) triggering the instability, see Fig.~\ref{FFT_grid_points}. 
The shaded regions comprise the particles subject to the FT (at
$x=-0.5$ and $x=0.5$). 
The maximum of the FT gives the dominant mode $k$ and its
corresponding velocity amplitude, which we compare with the analytical
model.

\label{lastpage}

\end{document}